\documentclass[times,twocolumn,final,nopreprintline]{elsarticle}
\usepackage{tabularx}
\usepackage{blkarray}
\usepackage{framed,multirow}

\usepackage{amssymb}
\usepackage{latexsym}

\usepackage{url}

\usepackage{hyperref}

\usepackage{caption}

\usepackage{booktabs}

\journal{Machine Learning with Applications}

\usepackage{amsmath}

\usepackage{float}

\usepackage{bbm}
\usepackage{bm}
\usepackage{appendix}
\usepackage{mathtools}
\usepackage{color}

\usepackage{subcaption}
\usepackage{natbib}

\usepackage{multirow}

\usepackage{hyperref}

\usepackage{float}

\usepackage{makecell}

\begin{document}

% \verso{Ozan Ciga \textit{et~al.}}

\begin{frontmatter}

\title{Self supervised contrastive learning for digital histopathology}%

\author[1]{Ozan Ciga}
%\author[1]{Ozan Ciga\corref{cor1}}
%\cortext[cor1]{Corresponding author: e-mail: ozan.ciga@mail.utoronto.ca}
\author[3]{Tony Xu}
\author[1,2]{Anne Louise Martel}

\address[1]{Department of Medical Biophysics, University of Toronto, Canada}
\address[3]{Department of Electrical and Computer Engineering, University of British Columbia, Canada}
\address[2]{Physical Sciences, Sunnybrook Research Institute, Toronto, Canada}

%\received{}
%\finalform{}
%\accepted{}
%\availableonline{}
%\communicated{}

\begin{abstract}
%%%
Unsupervised learning has been a long-standing goal of machine learning and is especially important for medical image analysis, where the learning can compensate for the scarcity of labeled datasets. A promising subclass of unsupervised learning is self-supervised learning, which aims to learn salient features using the raw input as the learning signal. In this paper, we use a contrastive self-supervised learning method called SimCLR that achieved state-of-the-art results on natural-scene images and apply this method to digital histopathology by collecting and pretraining on 57 histopathology datasets without any labels. We find that combining multiple multi-organ datasets with different types of staining and resolution properties improves the quality of the learned features. Furthermore, we find using more images for pretraining leads to a better performance in multiple downstream tasks. Linear classifiers trained on top of the learned features show that networks pretrained on digital histopathology datasets perform better than ImageNet pretrained networks, boosting task performances by more than 28\% in $F_1$ scores on average. These findings may also be useful when applying newer contrastive techniques to histopathology data. Pretrained PyTorch models are made publicly available at  \url{https://github.com/ozanciga/self-supervised-histopathology}.

%%%%
\end{abstract}

%\begin{keyword}
% \MSC 41A05\sep 41A10\sep 65D05\sep 65D17
%% Keywords
%\KWD Self supervised learning, digital histopathology, whole slide images, unsupervised learning
%\end{keyword}

\end{frontmatter}

%\linenumbers


\section{Introduction}

\begin{figure*}
         \centering
         \includegraphics[width=\textwidth]{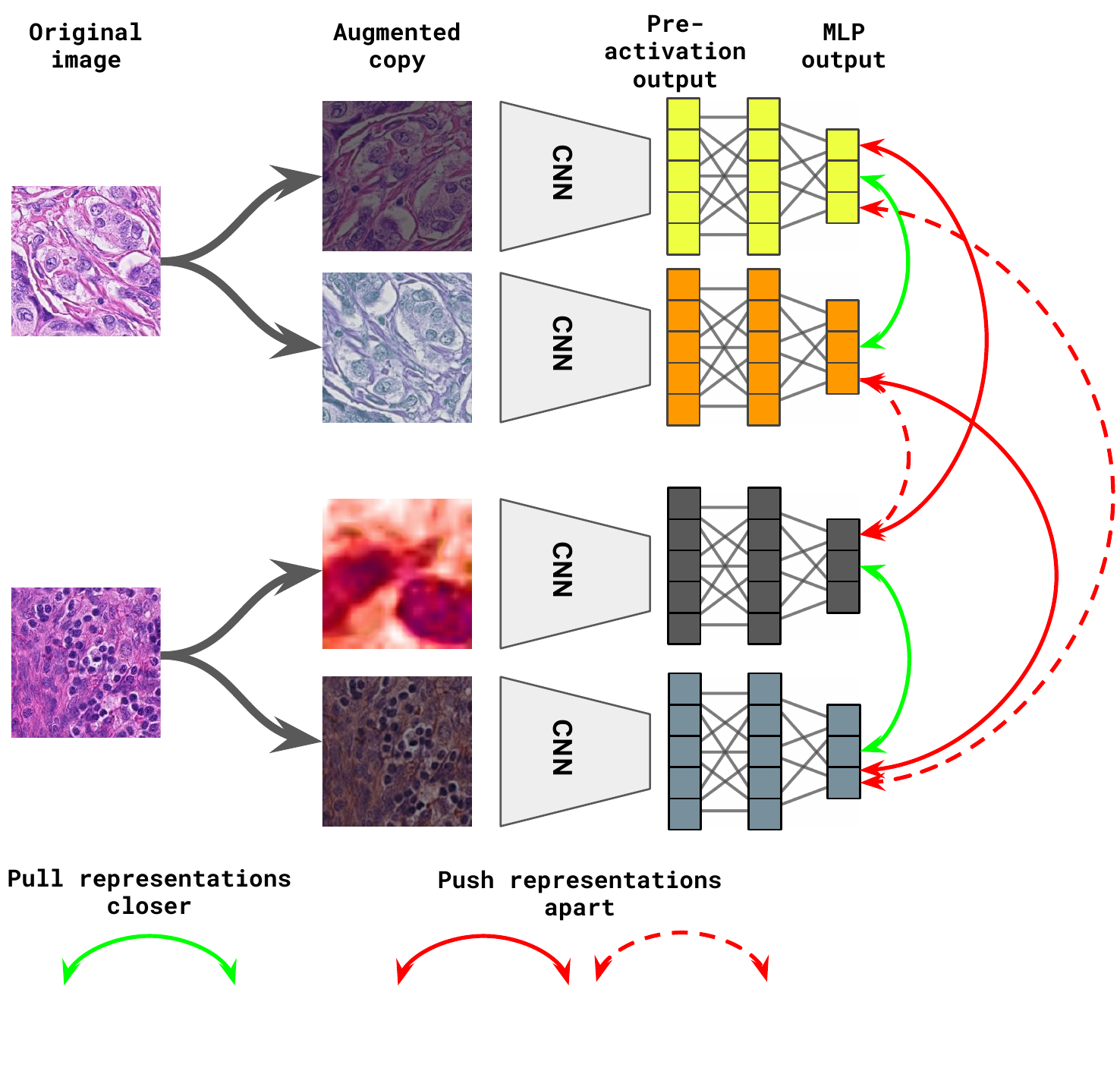}
     \caption{Overview of the method. The block represented as \textit{CNN} is identical for all image patches, and only replicated in the figure for clarity. We augment each image twice, extract features using the CNN at the pre-activation output, and transform these features into a lower embedding space at the multi-layer perceptron (MLP) output. While the augmented image patches from the same original source are pulled together (green arrows) in the feature space (at the MLP output), image patches from different sources are pushed apart using the NT-Xent loss function (both solid and dashed red arrows represent an identical push operation, shown as such in figure for clarity). The aim is to pretrain the network such that similarity between augmented views is used as a signal for learning salient features at the pre-activation layer. The figure is inspired from \cite{chen2020simple}.}\label{fig:proposed_architecture_full}
\end{figure*}

The number of labeled images in machine learning tasks is found to be positively correlated with the task performance; however, labeled data is scarce and expensive. The problem is exacerbated in medical image analysis tasks, where expert annotations are often required and crowdsourcing is not usually an option.  In many cases, labeling must also be done on-site due to regulations regarding dissemination of private patient data. In  any medical image analysis task, the most laborious and time-consuming step tends to be labeling the data and several approaches have been proposed to mitigate this \textit{data annotation bottleneck}. Unsupervised and self-supervised methods that can utilize unlabeled data, and semi-supervised methods that use partially labeled data, have been found to improve task performance \citep{peikari2018cluster, komura2018machine, campanella2019clinical}. 

% A common reason for overfitting to neural networks, especially in low data regimes, is feature memorization. Without any prior knowledge, the network quickly learns features that help to decrease the training loss but are not generalizable to an unseen test set. Pretraining aims to learn more general features which can be fine tuned for better performance. While most practitioners use ImageNet pretraining to mitigate overfitting, these features are learned on natural-scene images. Due to differences between such images and the histopathology domain, a certain fraction of these pretrained features are unlikely to be useful for histopathology tasks. This leads to capacity underuse. To mitigate these issues, self-supervised learning aims to learn features from raw data from the domain of interest without using any labels.

Until recently, most self-supervised techniques have relied on natural-scene image properties which are not applicable to histopathology images. However, recent contrastive learning approaches can be applied to digital pathology images (see Section \ref{sec:related_work}). In this work, we use residual networks pretrained with self-supervised learning to learn generalizable features. We employ SimCLR \citep{chen2020simple}, a contrastive self-supervised technique that has comparable performance to the supervised ResNet 50 network on top-1 classification accuracy for the ILSVRC2012 dataset. We pretrain residual networks with this method and use the pretrained networks in downstream tasks for multiple tasks on multiple, multi-organ digital histopathology datasets. We explore different data sampling strategies to understand the amount and the type of data that leads to a representation that improves task performances over ImageNet pretraining or training from scratch. Specifically, we examine the impact of the number of images used for pretraining, resolution, staining, and tissue type on the learned representations and downstream task performance. Furthermore, we compare multiple image augmentation strategies to identify the best practices when pretraining for digital histopathology. Our results indicate that pretraining with unlabeled histopathology images can improve task performances over Imagenet pretraining and mitigate labeled data requirements for various classification, regression, and segmentation tasks.

% We aim to determine the best practices in pretraining a network for digital histopathology. 

\section{Related work}\label{sec:related_work}

Unsupervised learning has been a long-standing goal of computer-aided diagnostic systems. Previously, sparse and variational autoencoders have been used for unsupervised nuclei detection and transfer learning (e.g., learning filter banks at multiple scales) \citep{xu2015stacked, chang2017unsupervised, hou2019sparse} and generative adversarial networks have been used for tissue and cell-level classification, nuclei segmentation, and cell counting \citep{hu2018unsupervised}. Most of these unsupervised methods are not applicable to structures larger than cells or to tasks more complicated than distinguishing tissue and cell types, mostly due to the small image sizes they are capable of working with (e.g., $64\times64$ pixels), and due to the limited information that can be encoded by such methods. They may also require custom networks, and may only be applicable to applications with specific resolution and staining properties. %However, these limitations are not unique to digital histopathology, where most of the fully unsupervised methods in machine learning has also limited applicability.  % Due to such limitations, unsupervised learning has not made a clinical impact.

Self-supervised learning is a promising subclass of unsupervised learning, where the raw input data is used to generate the learning signal instead of a prior such as mutual information or cluster density. These methods can generally be applied to larger images and work with standard architectures such as residual networks \citep{He2016}. Context-based self-supervised methods rearrange the image input and task the network to perform spatial reordering. For instance, \cite{noroozi2016unsupervised} tile an image into nine square pieces and then shuffle the tiles, whereas \cite{gidaris2018unsupervised} rotate the input image in $90^{\circ}$ angles. In either case, the aim is to obtain the original input using a neural network, which is effectively pretrained for downstream tasks by performing predictions on tile orderings or rotation angles, respectively. While contextual information can be exploited in natural scene images to obtain meaningful representations, structures within histopathological images are elastic and may form infinitely many valid groupings. Therefore, these techniques are not directly applicable to the histology domain. For instance, predicting rotations is not a viable task for whole slide images, since cells and surrounding structures will have a valid arrangement in the rotated image as well. In histopathology, \cite{gildenblat2019self} use spatial adjacency as a signal for similarity, and the pretraining task is to label image pairs as similar or dissimilar based on the spatial distance on a whole slide image. \cite{tellez2019neural} also utilize contrastive learning by applying augmentations on image patches extracted from WSIs. Their aim is to learn salient features by distinguishing if two augmented images are from the same source image.

% In addition to having only limited applicability for digital histopathology, context-based approaches are based on handcrafted pretext tasks, which may bias and limit the learning. More recently, contrastive approaches based on learning latent-space features by discriminating between unlabeled training samples achieved state-of-the-art results in computer vision tasks without relying on a heuristically modelled pretext task.

More recently, contrastive approaches based on learning latent-space features by discriminating between unlabeled training samples achieved state-of-the-art results in computer vision tasks. Such contrastive learning methods assume that under minor transformations, two views of the same image patch should have similar feature representations \citep{becker1992self}. Importantly, since contrastive methods only rely on consistency at the instance level, they do not require any spatial regularity between or within instances and are applicable to digital histopathology images. The consistency assumption has been exploited by \cite{dosovitskiy2014discriminative} to obtain a parametric feature representation for each training instance. Later, \cite{wu2018unsupervised} extended \cite{dosovitskiy2014discriminative} into a non-parametric feature representation using a dynamic memory bank to store latent features of data samples. The memory bank is used for selecting negative examples for each training instance, where any image that is not another view or augmentation of the original training instance is considered negative. The memory bank is then used to obtain negative samples without the need to recompute feature vectors. The use of simple image augmentations (e.g., resizing images, horizontal flips, color jittering, etc.) and memory banks have proved successful in learning representations by maximizing the mutual information between latent representations of positives \citep{bachman2019learning, henaff2019data, he2020momentum}. Data augmentations have also been utilized by applying the appropriate contrastive loss function on the feature vectors of positive and negative image pairs \citep{tian2019contrastive, misra2020self}. 

Recently, \cite{chen2020simple} proposed a contrastive learning approach that does not require a custom network or a memory bank, but instead relies on using a large number of minibatch instances ($\geq 256$) for obtaining negative samples per training instance. By doing so, they were able to improve the quality of learned representations by providing more negative samples per training instance over training epochs. Along with a few architectural improvements, this method outperforms aforementioned techniques by a large margin (+7\%) and has comparable performance to the supervised ResNet 50 network when its features are used in training a linear classifier for the ImageNet ILSVRC-2012 dataset \citep{ILSVRC15}. This method was later incrementally improved by \cite{grill2020bootstrap} through techniques such as exponential moving averaging of the model weights \citep{tarvainen2017mean} or by knowledge distillation and using larger projection layers \citep{chen2020big}. We choose not to explore these techniques in this paper as the added increase in performance ($\sim 4\%$ in ImageNet top-1 accuracy) cannot be justified by the increase in the number of trainable parameters (e.g., $\sim 800$ million trainable parameters versus $\sim 24$ million).

\section{The method}

\subsection{Contrastive self-supervised learning}\label{sec:method_simclr}

We use the contrastive learning method proposed by \cite{chen2020simple} which relies on maximizing agreement between representations of two stochastically augmented 
views of the same image (see Fig. \ref{fig:proposed_architecture_full}). Specifically, given an image $i$, a probabilistic augmentation function $f_{aug}(\cdot)$, a neural network or an encoder $f_{\theta}(\cdot)$ with parameters $\theta$ and an auxiliary projection layer $p_{\hat{\theta}}(\cdot)$ with parameters $\hat{\theta}$, the aim is to match $\ell_2$ normalized feature representations of two augmentations of the same image given by $\bm{z_i} = p_{\hat{\theta}}(f_{\theta}(f_{aug}(i)))$. Simultaneously, other images in a batch are made dissimilar from the image $i$ through a contrastive loss function called \textit{NT-Xent} (the normalized
temperature-scaled cross-entropy loss), defined as \begin{equation}
    \ell_{i,j}= -\log \frac{\exp(\textrm{similarity}(\bm{z_i},\bm{z_j})/\tau)}{\sum_{k=1}^{2N}\mathbbm{1}_{[k\neq i]}\exp(\textrm{similarity}(\bm{z_i},\bm{z_k})/\tau)}, 
\label{eqn:nt_xent}\end{equation} where $\tau$ is the temperature parameter that helps weigh different examples to achieve hard negative mining, $\mathbbm{1}$ is the indicator function which outputs 1 when $k\neq i$ and 0 otherwise, and the similarity function is a distance metric between two $\ell_2$ normalized vectors. We refer to output of $f_{\theta}(\cdot)$ as the pre-activation layer output, and the output of $p_{\hat{\theta}}(\cdot)$ as the MLP output in Fig. \ref{fig:proposed_architecture_full}, or $\bm{z_i}$. The auxiliary projection layer is a single hidden layer MLP which is used to project the pre-activation layer output into a lower embedding space. Comparing $\bm{z_i}$ and $\bm{z_j}$ was found to be more effective in learning representations than directly comparing the pre-activation layer outputs. For our experiments, we use cosine similarity defined as $\textrm{similarity}(\bm{u},\bm{v}) = \bm{u}^T \bm{v} / \left\lVert\bm{u}\right\rVert \left\lVert \bm{v}\right\rVert$. The authors experimentally find that NT-Xent helps learn better representations than similar loss functions such as margin \citep{schroff2015facenet} or logistic \citep{mikolov2013efficient} losses.

For each pretraining step with a batch size of $2N$, each augmented image has one similar (or positive) and $2(N-1)$ dissimilar (or negative) samples. By using samples in the same batch as negative samples, we avoid the expensive explicit negative example mining present in many methods \cite{wu2018unsupervised, he2020momentum} and are able to scale up batch sizes where each pretraining step simultaneously optimizes for $\sim 4N^2$ feature vectors.

\begin{figure*}
     \centering
     \begin{subfigure}[b]{0.45\textwidth}
         \centering
         \includegraphics[width=1\textwidth, height=240pt]{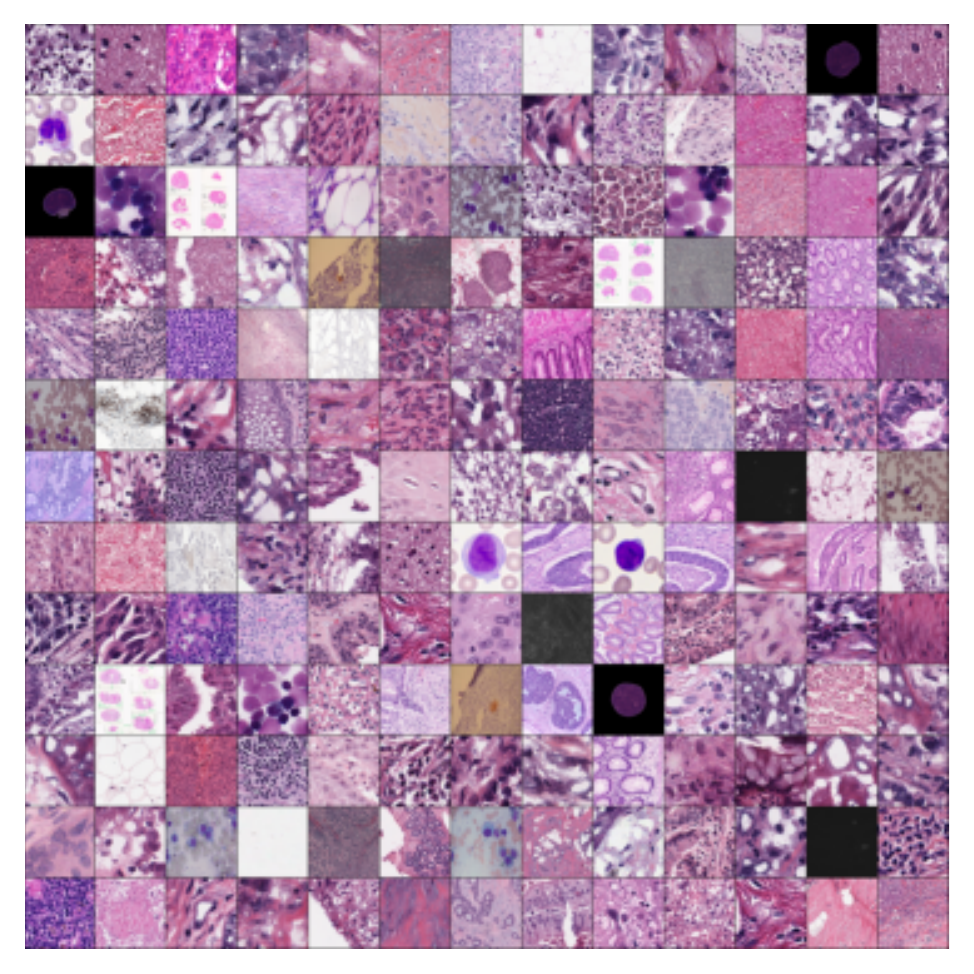}
         \caption{Multiple dataset sampling.}
         \label{fig:diversity_sampling_yes}
     \end{subfigure}
     \begin{subfigure}[b]{0.45\textwidth}
         \centering
         \includegraphics[width=1\textwidth, height=240pt]{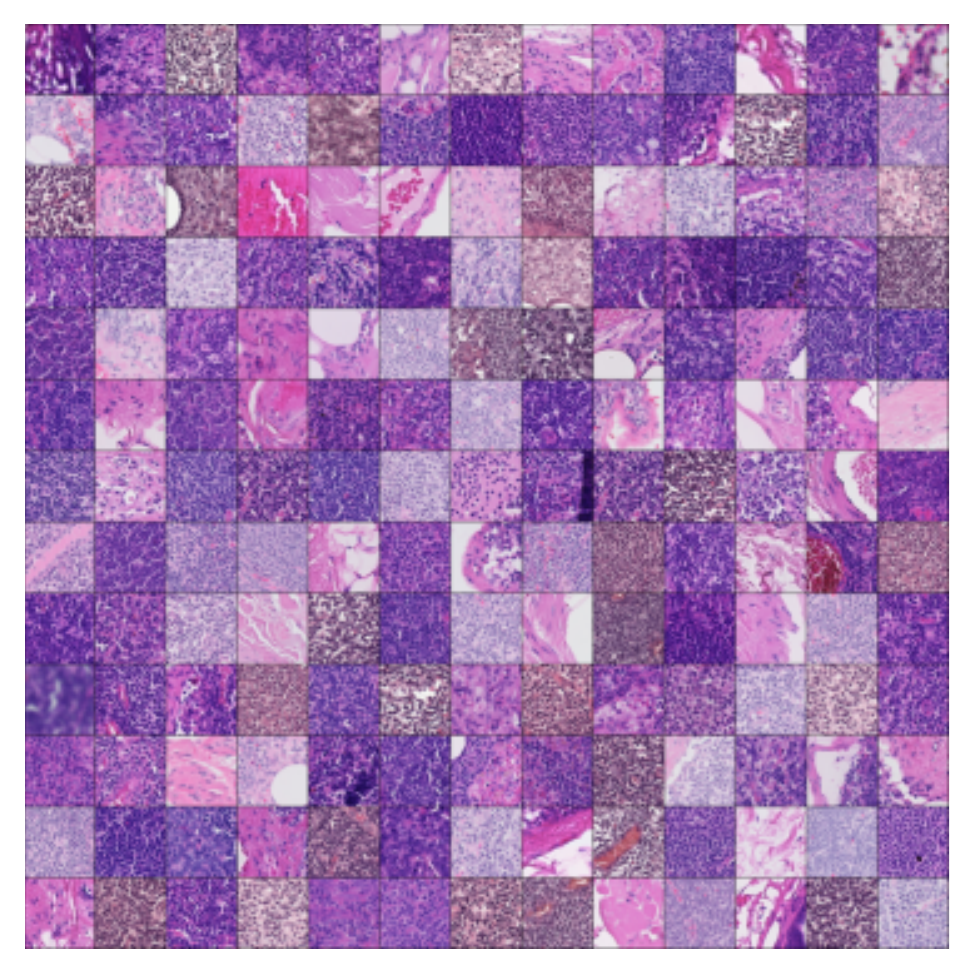}
         \caption{Sampling a single dataset (Camelyon 17).}
         \label{fig:diversity_sampling_no}
     \end{subfigure}
        \caption{Defining image diversity in the context of digital histopathology. Images evenly sampled from all 60 datasets visually look more diverse compared to sampling only from a single dataset containing multiple WSIs.}\label{fig:diversity_sampling}
\end{figure*}

\paragraph{Defining image diversity for histopathology} The contrastive method exploits the variability in visual properties between image patches to learn salient features. Obtaining visually diverse patches is challenging for digital histopathology, especially when images are viewed under high resolution. Given multiple unlabeled datasets, we assume that selecting images with different staining, resolution and tissue types will lead to a more diverse dataset compared to selecting image patches extracted from the same WSI or sampling images from the same dataset. In our early experiments, we found that pretraining with the dataset constructed using the former approach resulted in better validation performance than constructing a pretraining dataset where images are sampled from a single dataset. Furthermore, when images are inspected visually, the former approach exhibits more diversity (see Fig. \ref{fig:diversity_sampling}).

\section{Experiments}\label{sec:experiments}

\subsection{Pretraining datasets} Pretraining datasets include images from sites such as blood, breast, lymph, colon, bone, prostate, liver, pancreas, bladder, cervix, esophagus, head, neck, kidney, lung, thyroid, uterus, bone marrow, skin, brain, stomach, and ovary. Out of the total 57 datasets, 22 are comprised of image patches, 35 are WSI datasets. Most datasets are stained with hematoxylin and eosin (H\&E) and come at the highest resolution of $0.25\mu m/pixel$ (commonly referred to as $40\times$). While the majority of the WSI datasets are from The Cancer Genome Atlas Program (TCGA) and Clinical Proteomic Tumor Analysis Consortium (CPTAC) databases, we also use multiple public challenge datasets from lymph and breast cancer tasks. Image patch datasets are collected from various publicly available sources and exhibit similar organ diversity as WSI datasets. For the detailed list of pretraining datasets, please refer to \ref{apx:Pretraining datasets}.

\subsection{Validation experiments}

\paragraph{Validation datasets} 

We validate pretrained networks trained under multiple settings (e.g., training only with a single dataset or tissue type, different number of training images etc.) on five classification, two segmentation and one regression dataset. The classification datasets include BACH (four class breast cancer classification), Lymph (three class malignant lymph node cancer classification), BreakHisv1 (binary breast cancer classification), NCT-CRC-HE-100K (nine class colorectal cancer tissue classification), and Gleason2019 (five class prostate cancer classification). Segmentation datasets are BACH (four class breast cancer segmentation on WSIs) and DigestPath2019 (WSI segmentation of early-stage colon tumors into healthy and cancerous tissue). Our single regression dataset is the BreastPathQ, which involves assigning a percentage cancer cellularity score to a given image patch. See \ref{apx:Validation datasets} for the detailed explanation of each validation set. In the following, abbreviations NCT, Bpq, and Dp19 refer to NCT-CRC-HE100K, BreastPathQ, and DigestPath2019, respectively.

\paragraph{Tasks} 

We compare pretrained networks with randomly initialized and ImageNet pretrained Resnet 18, 34, 50 and 101. For segmentation tasks, we use a UNet architecture \citep{ronneberger2015u} which contains a pretrained encoder and a randomly initialized decoder. We compare two supervised training settings: fine-tuning and last layer training. In order to assess the learned representations directly, we freeze each residual network at the pre-activation layer and only train a linear classifier or a regressor on the learned representations (\textit{last layer} training). The last layer supervised training setting is omitted for the segmentation task, since the UNet-like decoder \citep{Yakubovskiy:2019} that we use contains a comparable number of trainable parameters to the pretrained Resnet encoder, and can mitigate the benefits of using a pretrained encoder or ImageNet initialization. We also train each network without freezing any layers (\textit{fine-tuning}). While fine-tuning is commonly employed, freezing various layers of a network may be used to avoid overfitting whenever the training dataset is small. Furthermore, pretrained features can also be used in clustering (see Section \ref{sec:pretrained_features}), feature selection, and in more traditional machine learning methods such as decision trees and support vector machines.

\paragraph{Validation setup} We train for 100 epochs per experiment, use Adam optimizer, a batch size of 128 with a weight decay of 0.00001 for Resnet18 models, and 0.0001 otherwise. We found that 100 epochs is enough for convergence for each validation dataset for all the networks (e.g., Resnet101) considered in this work. We use 50\% of the original dataset as the training, 25\% as validation, and 25\% as the test set. We use macro $F_1$ score as our validation metric for the classification and segmentation tasks. Macro F1 weighs each class equally regardless of the number of samples per class, which accounts for the class imbalance seen in most digital histopathology datasets. For the regression task, we compute the mean absolute error ($L_1$) on the predictions and the ground truth regression labels (varying between 0 to 100\%). The test metrics corresponding to the maximum validation metrics are reported. Please refer to the \ref{apx:eval_metrics} for the definitions of evaluation metrics.

\section{Results} 

Unless otherwise stated, all pretraining experiments are conducted with the Resnet18 model, and validation experiments were conducted without freezing any layers of the network. We use ``pretraining" to refer to the unsupervised training. Pretrained networks are then used for ``supervised" training on validation datasets. A dataset with half a million images of size $224\times224$ pixels can be pretrained for 1000 epochs using PyTorch on 4 Tesla P100 GPUs in about 24 hours for a Resnet 18 model. For a detailed explanation of augmentation and hyperparameter selection for the pretraining stage, please refer to \ref{apx:hyperparam_selection}.

In the following, we report the average classification and segmentation macro $F_1$ scores, as well as the $L_1$ error for the regression, in the appropriate subsection. The expanded results with individual datasets for each experiment are given in \ref{apx:detailed_results}.

\subsection{Overall comparison}\label{sec:overall_comparison}

We compare networks pretrained with self supervision, ImageNet initialization, and randomly initialized networks for Resnet 18, 34, 50, and 101 in Table \ref{tab:overall_comparison}. For comparison to other self- and unsupervised methods, please refer to \ref{apx:other_methods}. % For a visual comparison between features generated by self supervision and ImageNet initialization, please refer to \ref{apx:filter_activations}}.

\begin{table}[H]
\centering
   \caption{The downstream task performance of networks trained on top of the pretraining, ImageNet initialization, and randomly initialized network. We report $F_1$ scores (higher is better) averaged over five validation datasets for the classification task (Cls.), mean $L_1$ error difference (lower is better) between the ground truth and the predicted cellularity percentage for one dataset for regression (Reg.), and average two $F_1$ scores (higher is better) for segmentation (Seg.).}\label{tab:overall_comparison}

    \begin{tabular}{c c| c c c}
    Resnet & Pretraining & Cls. & Reg. & Seg.  \\
    \midrule
      18 & Random & 55.7 & 13.5 & 59. \\
       & Imagenet & 66.7 & 9.3 & 73.5 \\
       & Self supervised & 77.0 & 6.6 & 74.0 \\
    \midrule
      34 & Random & 57.7 & 13.7 & 58.7 \\
       & Imagenet & 65.7 & 7.1 & 66.2 \\
       & Self supervised & 77.9 & 5.9 & 66.6 \\
    \midrule
      50 & Random & 55.3 & 13.3 & 58.8 \\
       & Imagenet & 65.8 & 7.0 & 69.7 \\
       & Self supervised & 76.9 & 5.5 & 66.7 \\
    \midrule
      101 & Random & 53.0 & 15.1 & 57.7 \\
       & Imagenet & 67.0 & 6.7 & 69. \\
       & Self supervised & 76.2 & 5.3 & 65.9 \\
    \bottomrule
    \end{tabular}
\end{table}

We find that pretraining is superior to ImageNet on classification and regression tasks for all settings. Self supervision is comparable to ImageNet for segmentation in Resnet 18 and 34. For larger networks, ImageNet performs better.

\subsection{Pretraining is most useful when only a small number of training images is available}

\begin{figure*}
     \centering
     \begin{subfigure}[b]{1\textwidth}
         \centering
         \includegraphics[width=1\textwidth]{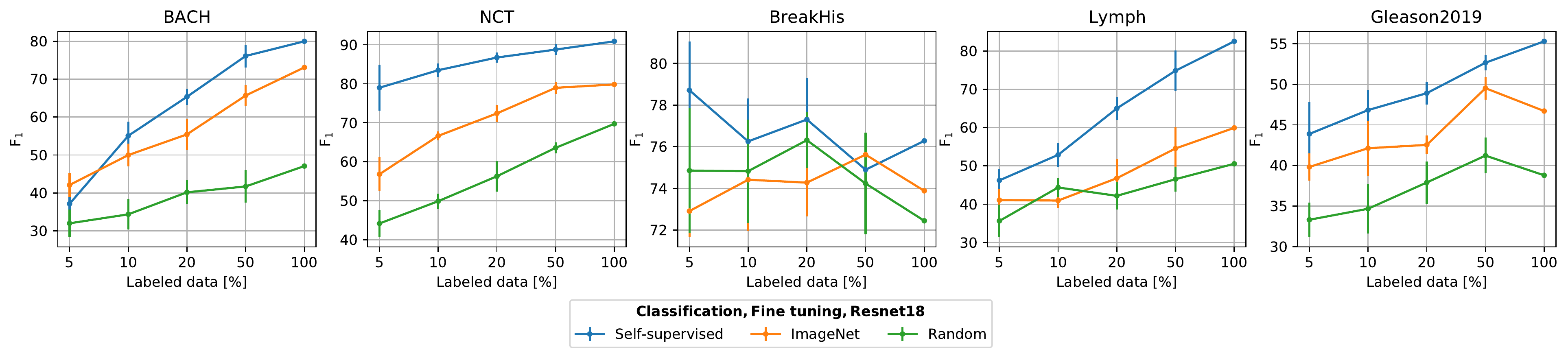}
     \end{subfigure}
     \vfill
     \begin{subfigure}[b]{1\textwidth}
         \centering
         \includegraphics[width=1\textwidth]{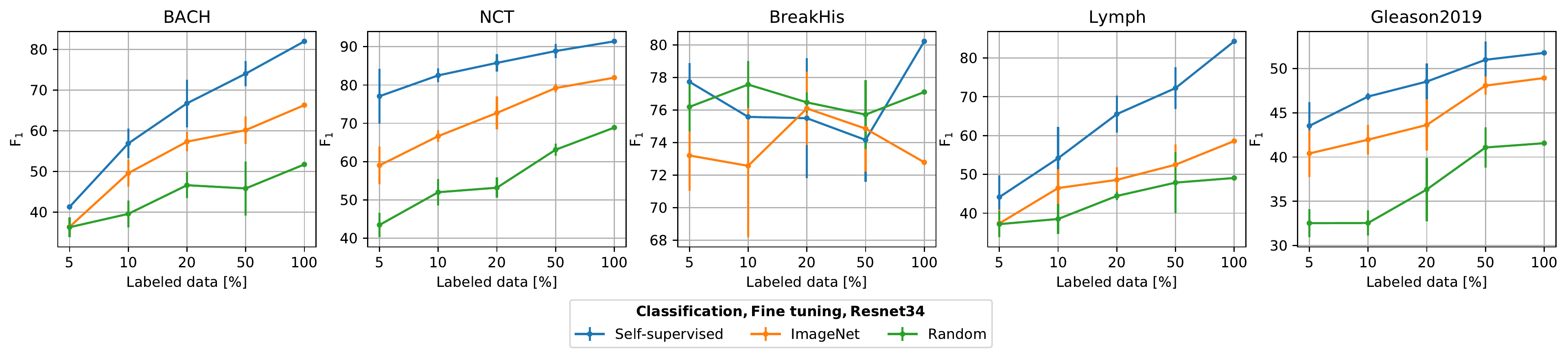}
     \end{subfigure}
     \vfill
     \begin{subfigure}[b]{1\textwidth}
         \centering
         \includegraphics[width=1\textwidth]{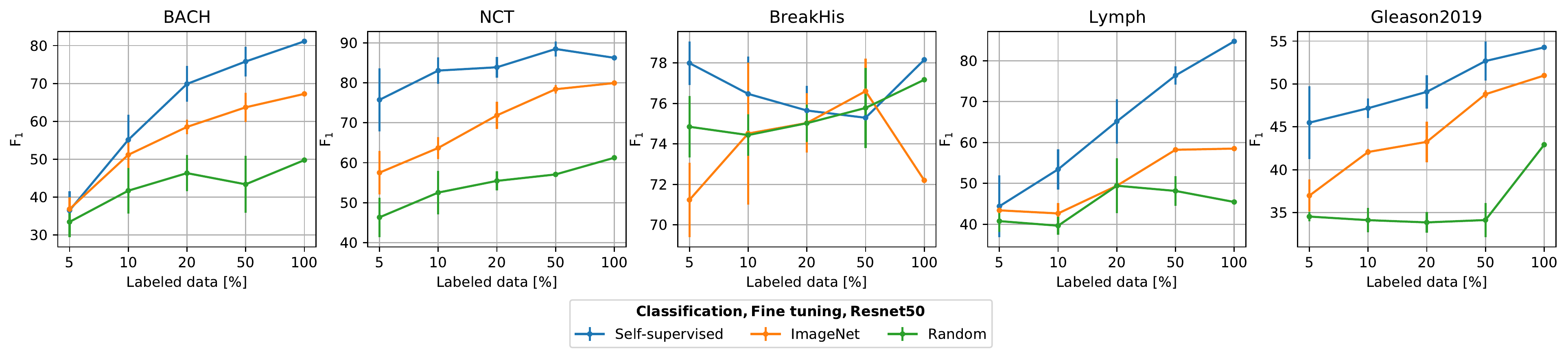}
     \end{subfigure}
     \vfill
     \begin{subfigure}[b]{1\textwidth}
         \centering
         \includegraphics[width=1\textwidth]{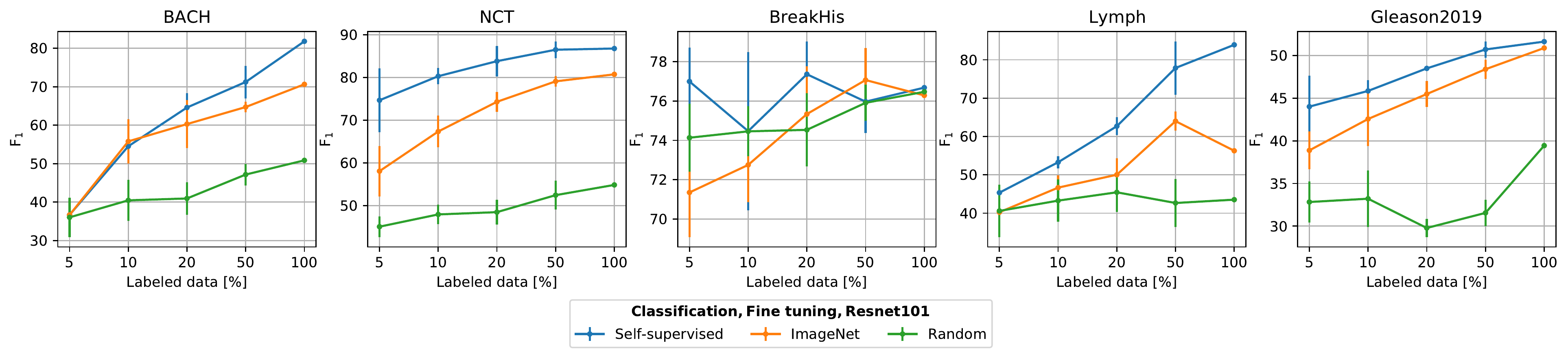}
     \end{subfigure}
     \vfill
      \begin{subfigure}[b]{1\textwidth}
     \centering
     \includegraphics[width=1\textwidth]{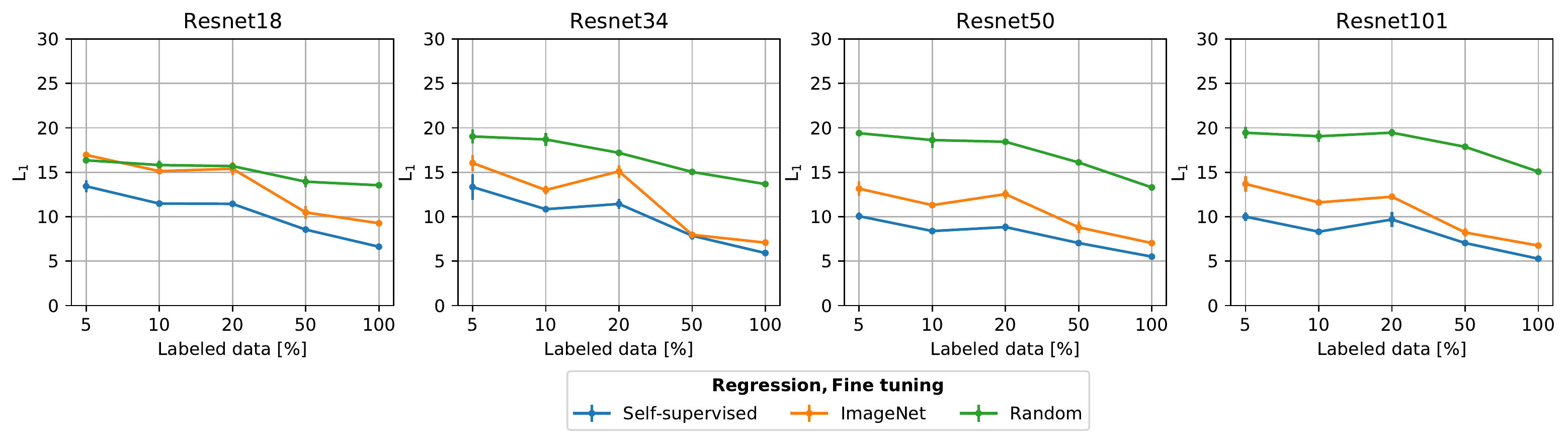}
 \end{subfigure}
        \caption{Classification results showing the macro $F_1$ score on the five validation datasets, and the regression results showing the mean absolute error ($L_1$) on the BreastPathQ dataset for four differently sized Resnet models, when the supervised training was conducted without freezing any layers (fine tuning). We report the $F_1$ score and the $L_1$ error when \{5, 10, 20, 50, 100\}\% of the labeled data is used for supervised training. Each percentage setting was run five times using different samples where the average and the standard deviation of five runs are shown as dots and error bars.}\label{fig:percentage_exps_finetuning}
\end{figure*}

It is generally accepted that pretrained networks boost the performance in medical image analysis \citep{tajbakhsh2016convolutional} and digital histopathology tasks \citep{mehra2018breast}. We conduct validation experiments on the classification and regression tasks for all datasets using \{5, 10, 20, 50, 100\}\% of each training dataset, where for each percentage value we repeat experiments five times to obtain a better estimate of the performance. By limiting the amount of training data, we compare the effect of training dataset size on different pretraining settings. Specifically, we compare random initialization, ImageNet pretraining and self-supervised pretraining for networks Resnet 18, 34, 50 and 101. We report detailed results per dataset on Figure \ref{fig:percentage_exps_finetuning}.

We found that in the absence of a large training set, training a classifier or a regression network using a pretrained initialization can improve the performance. For instance, self supervision outperforms training from scratch (random setting) by over 40\% for the NCT dataset when only 5\% of the labeled images are used for training. When more data is used, self supervision still remains superior to both ImageNet and random initializations.

\subsection{Features obtained by pretraining are more representative for histopathological image patches than ImageNet features}\label{sec:pretrained_features}

\begin{figure*}
     \centering
     \begin{subfigure}[b]{1\textwidth}
         \centering
         \includegraphics[width=1\textwidth]{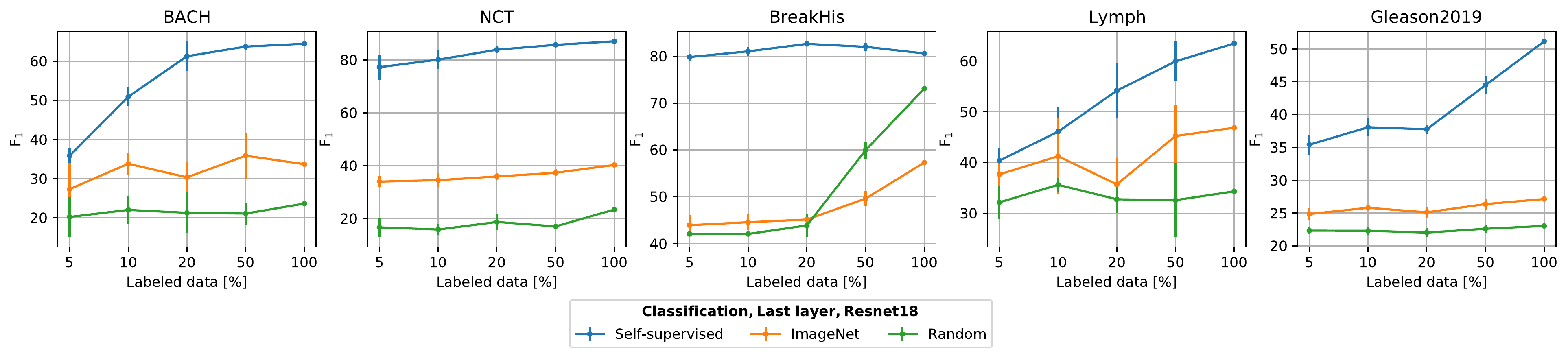}
     \end{subfigure}
     \vfill
     \begin{subfigure}[b]{1\textwidth}
         \centering
         \includegraphics[width=1\textwidth]{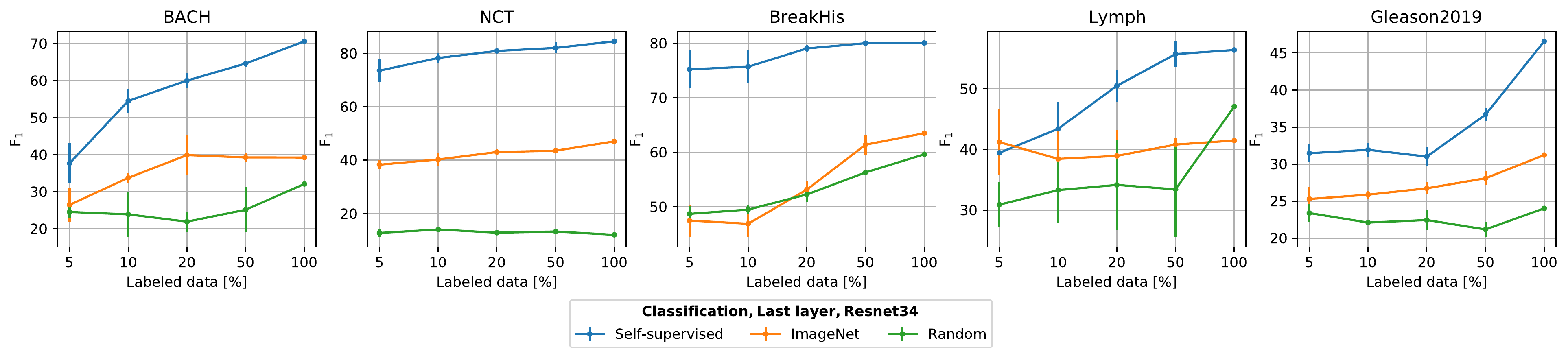}
     \end{subfigure}
     \vfill
     \begin{subfigure}[b]{1\textwidth}
         \centering
         \includegraphics[width=1\textwidth]{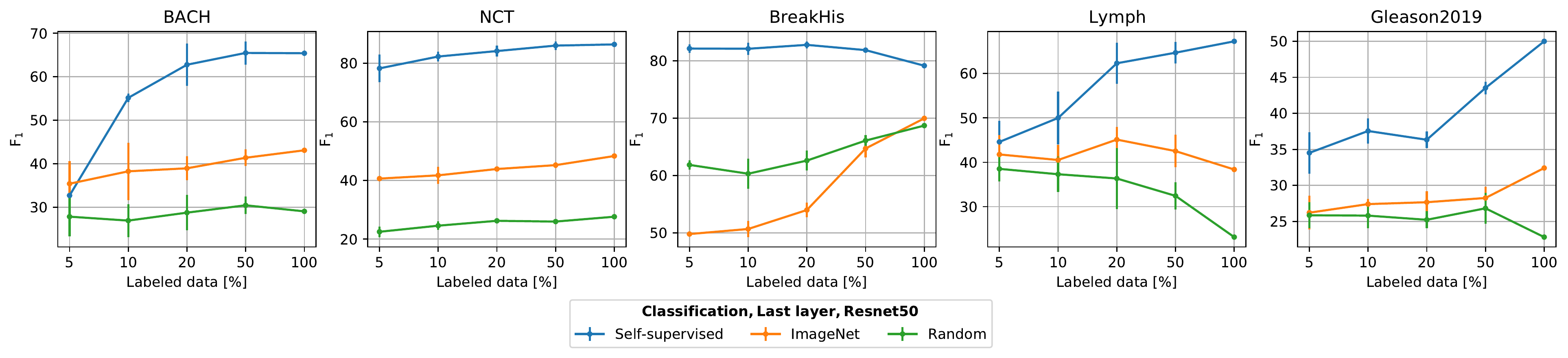}
     \end{subfigure}
     \vfill
     \begin{subfigure}[b]{1\textwidth}
         \centering
         \includegraphics[width=1\textwidth]{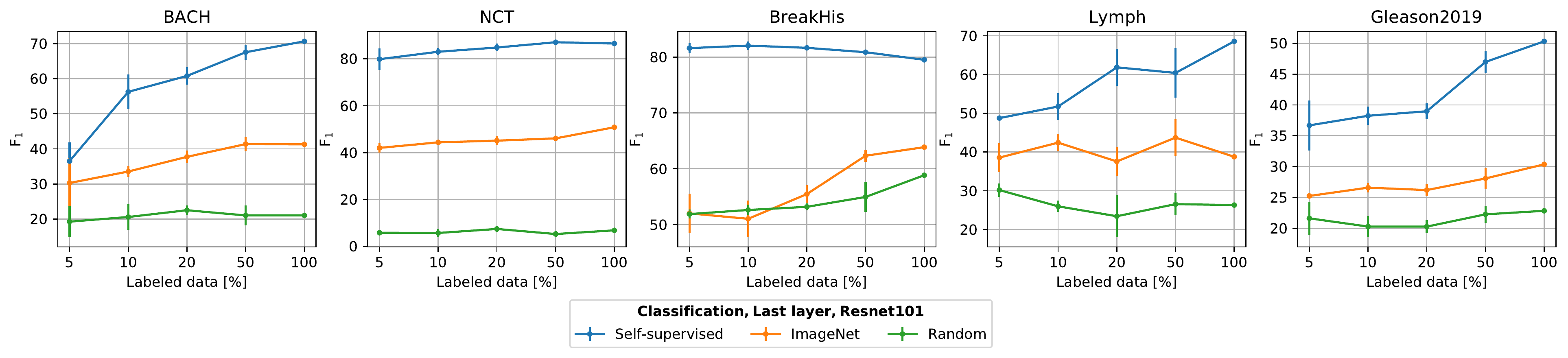}
     \end{subfigure}
     \vfill
      \begin{subfigure}[b]{1\textwidth}
     \centering
     \includegraphics[width=1\textwidth]{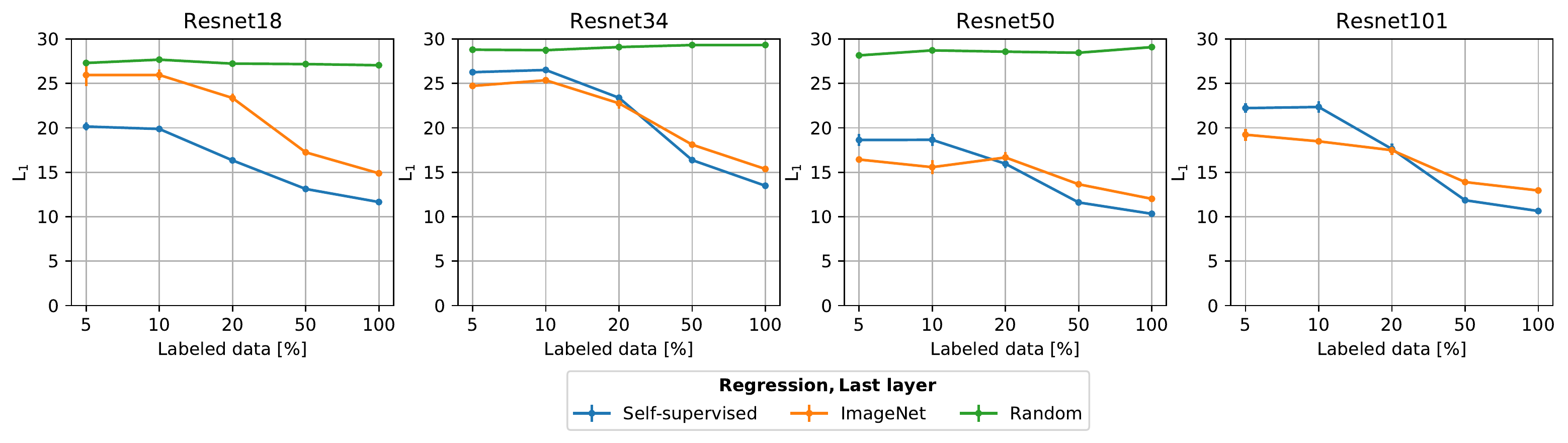}
 \end{subfigure}
        \caption{Classification results showing the macro $F_1$ score on the five validation datasets, and the regression results showing the mean absolute error ($L_1$) on the BreastPathQ dataset for four differently sized Resnet models, when the supervised training was conducted by freezing all the layers except the linear classification or the linear regression layer (last layer setting). We report the $F_1$ score when \{5, 10, 20, 50, 100\}\% of the labeled data is used for supervised training. Each percentage setting was run five times using different samples where the average and the standard deviation of five runs are shown as dots and error bars. For the Resnet 101, last layer training, randomly initialized network’s error is too large to fit in the plot.}\label{fig:percentage_exps_lastlayer}
\end{figure*}

We compare representations learned through self-supervision with the pretrained ImageNet features and randomly initialized residual network weights by training linear classifiers on the pre-activation layer of the Resnet models for the validation datasets. For Resnet 18 and Resnet 34, this amounts to 512 features, and to 2048 features for Resnet 50 and Resnet 101. We use 100\% of the available data for each experiment. We performed the same comparison for the regression task, where we froze the network at the pre-activation layer and trained a single regression layer. We do not include analogous frozen encoder experiments for the segmentation task since a decoder can contain millions of parameters that can be trained to achieve satisfactory performance regardless of the encoder weights. We report the average results in Table \ref{tab:last_layer_exps}, and the detailed results per dataset on Figure \ref{fig:percentage_exps_lastlayer}.

\begin{table}[H]
\centering
   \caption{The downstream task performance of linear classifiers trained on top of the learned features by self supervision, ImageNet initialization, and randomly initialized network. The network weights up to the pre-activation layer are frozen and not updated except for the linear classification layer. We report $F_1$ scores averaged over five validation datasets for the classification task, mean $L_1$ difference between the ground truth and the predicted cellularity percentage for one dataset for regression. Segmentation results using frozen encoders are omitted since a decoder can contain millions of parameters that can be trained to achieve satisfactory performance regardless of the encoder weights.}\label{tab:last_layer_exps}
    \begin{tabular}{ c c| c c c}
    Resnet & Pretraining & Classification & Regression \\
    \midrule
      18 & Random & 35.5 & 27.0 \\
       & Imagenet & 41.1 & 14.9 \\
       & Self supervised & 69.3 & 11.6 \\
    \midrule
      34 & Random & 35.0 & 29.3 \\
       & Imagenet & 44.5 & 15.4 \\
       & Self supervised & 67.6 & 13.5 \\
    \midrule
      50 & Random & 34.3 & 29.1 \\
       & Imagenet & 46.4 & 12.0 \\
       & Self supervised & 69.6 & 10.3 \\
    \midrule
      101 & Random & 27.2 & 2490.7 \\
       & Imagenet & 45.0 & 12.9 \\
       & Self supervised & 71.1 & 10.6  \\
    \bottomrule
    \end{tabular}
\end{table}

The self-supervised network achieves better results than ImageNet initialization, which indicates our method has learned domain-specific features that can be useful, especially when training samples are scarce. One may freeze and obtain features from various layers of the network for training various machine learning models, including neural networks, support vector machines or random forest classifiers.

\paragraph{Unsupervised clustering using the learned representations}

Learned representations can be directly used for clustering without any labeled training examples. In this section, we use learned representations to cluster image patches extracted from WSIs. Note that the learned features can also be used for querying an image to its nearest neighbors without clustering the dataset, which is useful in applications such as active learning for sample selection and various data retrieval systems.

\begin{figure*}
     \centering
     \begin{subfigure}[b]{0.45\textwidth}
         \centering
         \includegraphics[width=1\textwidth, height=240pt]{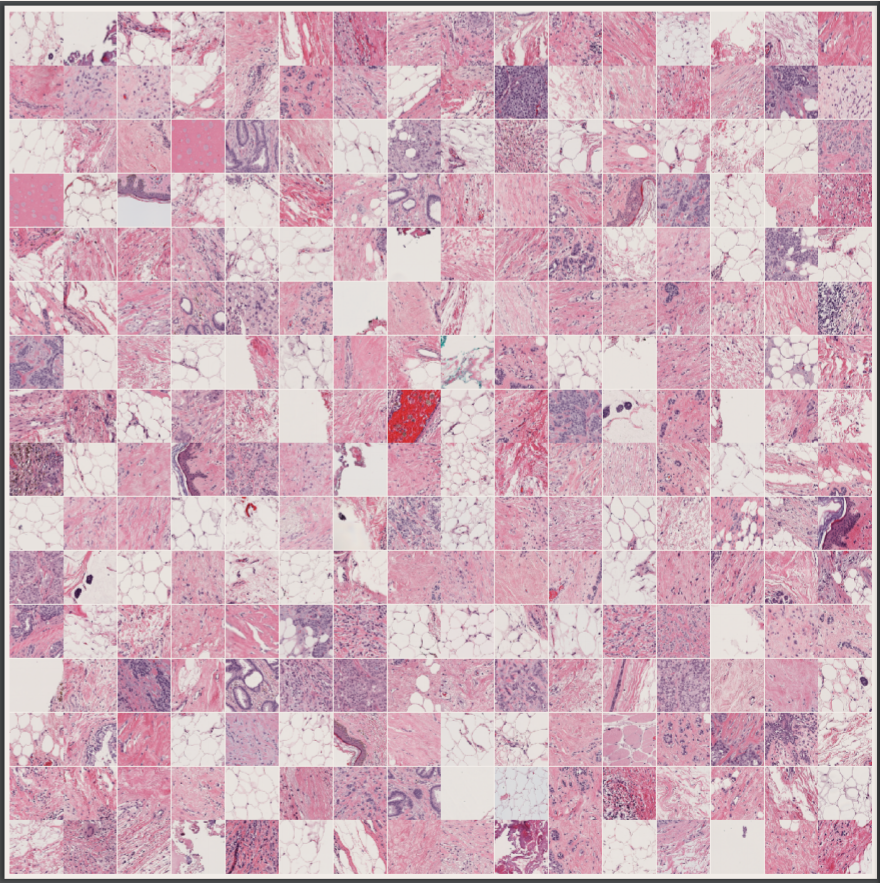}
         \caption{Random sampling}
         \label{fig:neg_sampling_random}
     \end{subfigure}
     \begin{subfigure}[b]{0.45\textwidth}
         \centering
         \includegraphics[width=1\textwidth, height=240pt]{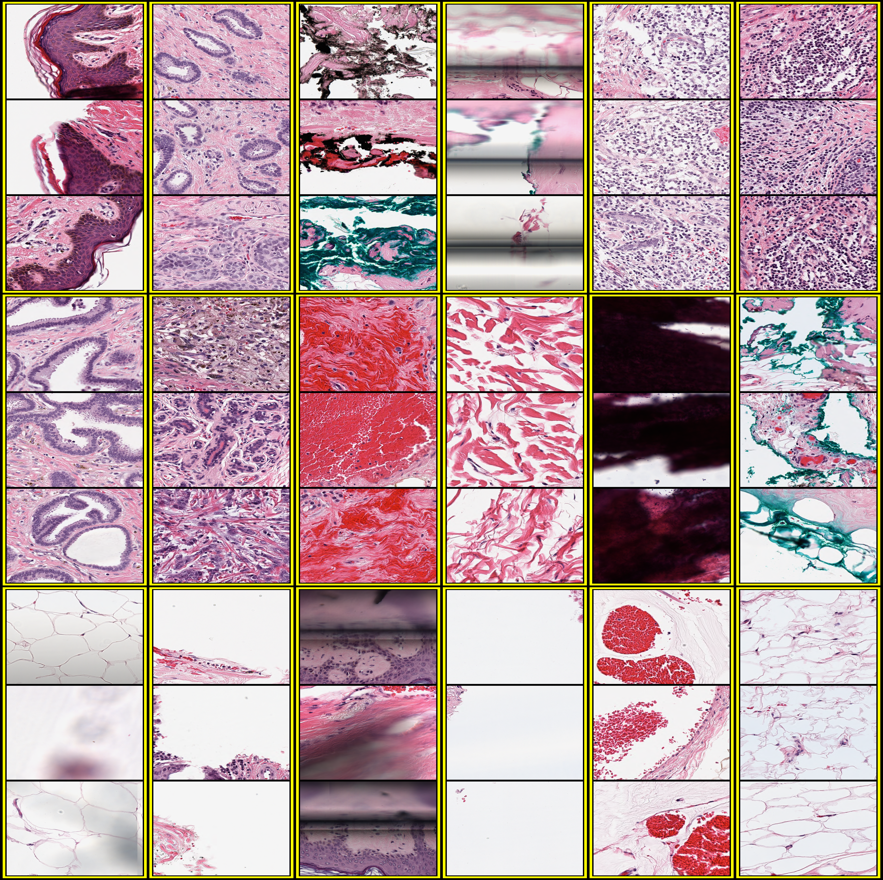}
         \caption{K-means sampling}
         \label{fig:neg_sampling_kmeans}
     \end{subfigure}
        \caption{Sampling image patches from whole-slide images randomly versus feature based K-means clustering. Each figure is extracted from an actual experiment. In Fig. \ref{fig:neg_sampling_kmeans}, each yellow box shows three samples closest to the center of a randomly selected cluster for illustration. Note that visually similar patches with features such as out-of-focus regions, creases, ink marks of \textit{different colors}, morphological structures such as ducts, outlines of the nipple, and red blood cells are clustered together. Furthermore, patches with varying nuclei formation patterns are also clustered roughly according to their density.}\label{fig:neg_sampling}
\end{figure*}

\begin{figure*}
      \centering
    \begin{subfigure}[b]{0.49\textwidth}
          \includegraphics[width=1\textwidth, page=1]{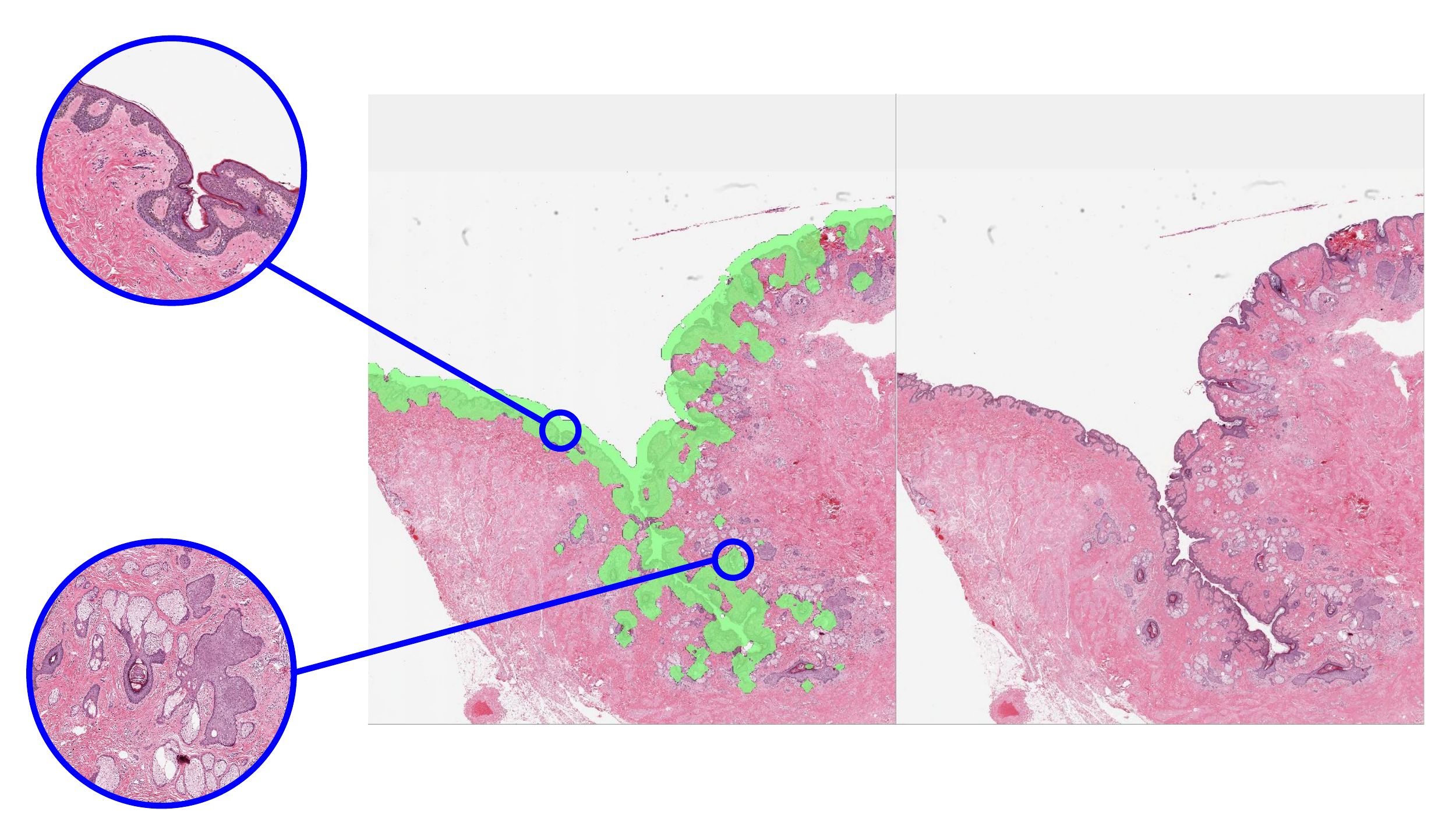}
          \caption{Squamous regions are highlighted.}
          \label{fig:neg_sampling_squamous}
    \end{subfigure}
     \begin{subfigure}[b]{0.49\textwidth}
         \centering
          \includegraphics[width=1\textwidth, page=2]{figures/unsup_cluster/Self_sup_unsupervised_image_samples.pdf}
          \caption{Clusters representing creases, out of focus regions, and smears are highlighted.}
          \label{fig:neg_sampling_crease}
     \end{subfigure}
     \caption{A practical application of unsupervised clustering in digital histopathology. We first cluster 1.4 million feature vectors corresponding to image patches into three thousand clusters without any supervision. Then, these clusters are used to highlight different regions of interest on whole-slide images.}\label{fig:sample_cluster_results}
 \end{figure*}

Negative mining is an important task in most tasks involving WSIs. For illustration of the saliency of learned representations, we perform negative mining on WSIs from a dataset including post-neoadjuvant therapy BRCA specimens with annotations of regions containing tumor \citep{peikari2017, Martel2019tcia}. Randomly sampling patches from a WSI will result in a large class imbalance in favor of negatives, which leads to an increased ratio of false negatives. One may mitigate this issue by aggregating the sampled patches and selecting a subset of ``representative" patches. This is done by clustering, where we rely on the perceived visual or morphological similarity of patches according to their relative distance to each other in the feature space. Specifically, we sample 1.4 million images from 69 WSIs and cluster images into three thousand clusters, where the number of clusters are determined using the Elbow heuristic \citep{ciga2021overcoming} (using the cluster sizes \{1000, 1500, ..., 10000\}, explained variance is 67.1\% when 2500 clusters is used, 83.3\% when 3000 clusters is used, and 91\% when 5000 clusters is used), using their feature representations ($\in \mathbb{R}^{512}$) generated by the Resnet18 trained using self supervision. The features are clustered using the mini-batch K-means algorithm. The resulting clusters can be seen in Fig. \ref{fig:neg_sampling} with three samples per cluster. In addition, we select a few clusters and highlight them on WSIs to segment various regions of interests without any supervision in Figure \ref{fig:sample_cluster_results}. %, and \ref{fig:neg_sampling_filledduct}.

\subsection{Using more unlabeled images in pretraining improves the downstream task performance}

Extracting patches of size $224\times224$ pixels from WSIs that contain foreground (refer to \ref{apx:Foreground filtering}) can lead to a few million patches per dataset, which can quickly become intractable. Therefore, we randomly sample a maximum of 100 patches from each WSI and use all images from non-WSI datasets to generate the unsupervised pretraining data. We use the maximum available resolution per WSI, which is $\sim 0.25\mu \text{m/pixel}$ or $\sim 0.50\mu \text{m/pixel}$ depending on the dataset. We use 206 thousand images from 23 non-WSI datasets and sample over four million patches from 35 WSI datasets containing around 25 thousand WSIs. We compare using $\{0.01, 0.1, 1, 10\}$\% of the sampled datasets with a maximum of two thousand and a minimum of 10 images per dataset if the number of images corresponding to the percentage setting for a dataset is below 10. In the end, we obtain 4, 40 and 400 thousand images per percentage setting. In addition, we pretrain a model using only 10 images per dataset ($\sim 0.01\%$ of all available images). In this setting, we sample 10 image patches from each WSI dataset as well, which can contain a few \textit{million} image patches across all WSIs in that dataset. Therefore, the dataset is drastically undersampled. The results can be seen in Table \ref{tab:dataset_size}.

\begin{table}[H]
\centering
   \caption{The effect of the number of images used for pretraining in the downstream task performance for classification, regression, and segmentation tasks. We report $F_1$ scores averaged over five validation datasets for the classification task, mean $L_1$ difference between the ground truth and the predicted cellularity percentage for one dataset for regression, and average two $F_1$ scores for segmentation.}\label{tab:dataset_size}
    \begin{tabular}{ c| c c c}
    \% of images & Classification & Reg. & Segmentation  \\
    \midrule
      $0.01\%$ & 66.5 & 8.5 & 62.7 \\
      $0.1\%$ & 73.9 & 7.8 & 72.8 \\
      $1\%$ & 77.0 & 6.8 & 70.8 \\
      $10\%$ & 79.1 & 6.6 & 73.5 \\
    \bottomrule
    \end{tabular}
\end{table}

We found that using a larger training set outperforms using a training set with fewer images. Notably, we observe diminishing returns with every order of magnitude of increase in dataset size. While the difference between using 0.01\% and 0.1\% of the available images is 7.4\% in average $F_1$ for the classification task, the difference between 1\% and 10\% is around 2.1\%.

% Specifically, we found that using $0.1\%$ vs. $1\%$ was comparable among different tasks. Using $10\%$ of the images was outperformed by all settings by a large margin for regression and segmentation tasks, and was on par with the 10 samples per dataset setting on classification tasks. 

% The degradation in performance when a large number of images are used is likely due to oversampling; if one region of tissue is oversampled resulting in a large number of very similar appearing patches, then the network may rely on features that are not generalizable in order to separate them. We find that for histopathological images, a moderately sized training set with a diverse set of images is crucial for obtaining useful representations.

\subsection{The impact of resolution on the pretraining}\label{sec:resolution_exps}

To assess the impact of resolution on the learned features, we pretrain four networks using images at different resolutions. We use a breast image dataset, originally aimed to identify the invasive ductal carcinoma grade at multiple resolutions \citep{BOLHASANI2020100341}. In the results shown in Table \ref{tab:resolution_exps}, $10\times$, $20\times$, and $40\times$ refer to $1\mu m/pixel$, $0.50\mu m/pixel$, and $0.25\mu m/pixel$, respectively. We pretrain networks using images with resolutions $10\times$, $20\times$, and $40\times$, in addition to using all available images ($10, 20, 40\times$).

\begin{table}[H]
\centering
   \caption{Examining the impact of resolution on task performance. We use images from one resolution for pretraining and evaluate the pretrained network on all available validation datasets. $10\times$, $20\times$, and $40\times$ refer to $1\mu m/pixel$, $0.50\mu m/pixel$, and $0.25\mu m/pixel$, respectively. We report $F_1$ scores averaged over five validation datasets for the classification task, mean $L_1$ difference between the ground truth and the predicted cellularity percentage for one dataset for regression, and average two $F_1$ scores for segmentation.}\label{tab:resolution_exps}

    \begin{tabular}{ c| c c c}
    Resolution & Classification & Regression & Segmentation  \\
    \midrule
      $10\times$ & 61.0 & 10.9 & 56.5 \\
      $20\times$ & 60.5 & 9.9 & 57.6 \\
      $40\times$ & 65.4 & 9.4 & 62 \\
      $10, 20, 40\times$ & 67.0 & 9.3 & 63. \\
    \bottomrule
    \end{tabular}
\end{table}

We find that the networks pretrained on images at higher resolutions tend to perform better in downstream tasks. Furthermore, we find that combining multiple resolutions further improves task performances.

\subsection{Transferability of features between tissue types and staining}\label{sec:transferability_of_features}

To assess if it is better to pretrain using datasets drawn exclusively from a similar tissue type to that present in the target task, we conducted separate experiments using Resnet 18 with three different tissue types: breast, lymph nodes and prostate. Specifically, we trained a network using only Camelyon 16 and Camelyon 17 datasets for lymph nodes, and TCGA-PRAD and Prostate-MRI datasets for prostate. Due to the greater availability of data for breast, we trained on a larger number of datasets: TCGA-BRCA, TUPAC16, TNBC, several datasets from Andrew Janowczyk, ICPR2014 and ICPR2012 (see Table \ref{tab:pretraining_datasets} for dataset descriptions). For each experiment, roughly the same number of training images were used ($\sim$ four thousand). The results can be viewed in Table \ref{tab:organ_exps}.

\begin{table*}
\centering
\begin{tabular}{c|c c c c c c|c|c c c}
\multirow{2}{*}{Tissue} & \multicolumn{6}{c|}{Classification} &
    \multicolumn{1}{c|}{Regression} &
    \multicolumn{3}{c}{Segmentation}\\
  & BACH & NCT & BreakHis & Lymph & Gleason2019 & Average & Bpq & BACH & Dp19 & Average \\
\hline
 Breast & 75. & 86. & 79.1 & 73.8 &  47.3 & 72.2 & 7.5 & 49.0 & 84.9 & 67.0 \\
 Lymph & 75.1 & 83.1 & 76. &  70.6 & 48.4 & 70.6 & 8.2 & 48.0 & 83.3 & 65.7 \\
 Prostate & 78.3 & 80.6 & 75.4 & 65.9 & 54.7 & 71.0 & 8.7 & 56.5 & 84.5 & 70.5 \\
 \bottomrule
\end{tabular}
\caption{Examining the transferability of features between organs. We use images from one organ for pretraining and validate it on all available validation datasets. We report $F_1$ scores averaged over five validation datasets for the classification task, mean $L_1$ difference between the ground truth and the predicted cellularity percentage for one dataset for regression, and average two $F_1$ scores for segmentation.}\label{tab:organ_exps}
\end{table*}

Despite the site-specific pretraining (i.e., the tissue used in pretraining), we did not observe a strong correlation between the pretrained model and the validation performance. For instance, prostate models outperformed breast models on the BACH dataset for both breast cancer classification and segmentation. Similarly, the lymph model was outperformed by the breast model on the malignant lymph cancer detection. However, it should be noted that for some datasets the site-specific pretraining yielded better performance. For instance, breast model performed the best on the BreastPathQ dataset, and prostate model outperformed other models on the Gleason prostate cancer grading. In addition, we observed poorer performance compared to training with all datasets, where the results in tables \ref{tab:overall_comparison} and \ref{tab_apx:overall_comparison} indicate using images from a more diverse pretraining dataset with comparable number of pretraining images leads to better performance across all tasks.

\section{Discussion}

In Section \ref{sec:overall_comparison}, we have observed that the self supervision performs better than the ImageNet initialization for the \textit{segmentation task} for Resnet 18 and 34. In contrast, for Resnet 50 and 101, ImageNet performs better. We hypothesize this is due to the increased number of trainable parameters for the decoder, diminishing the effect of pretraining the encoder.

In Section \ref{sec:transferability_of_features}, we have observed that the tissue type used as data for pretraining was not correlated to the downstream task performance on the validation datasets. We believe the lack of correlation and the degradation compared to using all tissue types can be due to the following factors: (1) limitations of the contrastive approach where representations are incapable of encoding domain-specific information in the absence of the \textit{other tissue types}, (2) convolutional networks are highly sensitive to visual properties such as staining, resolution and morphological shapes, and do not encode abstract features in the absence of a specific objective (e.g., cancer grading), (3) the network can only be \textit{incentivized} to encode a richer representation given a diverse pretraining dataset.

\section{Conclusion}

Our main objective in this work was to show that, by pretraining, we can learn better features to improve performances on multiple downstream tasks, including classification, regression, and segmentation. The self-supervised method outlined in this paper is the first method to consistently have comparable performance to ImageNet pretraining without additional complexity. To our knowledge, there is no prior research on histopathological image analysis with a training regimen that consistently reaches or surpasses supervised training. In addition, this is the first study which uses a very large number of images in digital histopathology setting: 23 image datasets with over 206 thousand patches and $\sim$ 25 thousand gigaresolution images in 35 datasets that consist of whole-slide images. 

We have shown that the success of the contrastive pretraining method heavily relies on the diversity of the unlabeled training set, as opposed to the number of images. This is an important consideration when one adopts a technique from the computer vision community, where most methods are validated on natural scene images that contain significantly more diversity than medical images. Furthermore, we have shown that the site which the training images were extracted from did not have a substantial effect on the quality of learned representations, as shown in Section \ref{sec:transferability_of_features}. While this shows a clear divergence from the training of human experts who focus on a specific organ, it also can significantly increase the number of available datasets in training such systems.

In this work, we focused on the simplest contrastive method which significantly improved the state-of-the-art and experimented under multiple settings outlined in Section \ref{sec:experiments} to understand the capabilities and limitations of the contrastive training for histopathology images. We believe that so long as the fundamentals of the sample contrastive training framework remain the same, the insights we obtained will still be valid for future work on self-supervision. Overall, we found that contrasting images which are visually distinguishable helped in learning salient representations. In contrast, images which look similar with small nuances that are important in histopathology (e.g., single cell tumors that only occupy a small portion of a given patch) were not suitable for contrastive learning, and led to noisy representations. As this is rarely the case for the natural-scene images, researchers working on digital histopathology images need to address domain-specific issues to bridge the gap between histopathology and computer vision in self-supervised learning.

\section*{Conflict of interest} \begin{flushleft} We have no conflict of interest to declare. \end{flushleft}

\section*{Acknowledgments} \begin{flushleft} This work was funded by Canadian Cancer Society (grant \#705772) and NSERC. It was also enabled in part by support provided by Compute Canada (www.computecanada.ca). The results shown here are in part based upon data generated by the The Cancer Genome Atlas (TCGA) Research Network: https://www.cancer.gov/tcga, and the Clinical Proteomic Tumor Analysis Consortium (CPTAC): https://cptac-data-portal.georgetown.edu/cptacPublic/. \end{flushleft}

\flushbottom

%%Harvard
\bibliographystyle{model2-names.bst}\biboptions{authoryear}
\bibliography{refs}

\clearpage
% \onecolumn

\appendix

\newpage

\section{Evaluation metrics}\label{apx:eval_metrics}

\begin{align}
  % &\frac{tp+tn}{tp+tn+fp+fn}&&\text{2-class accuracy}\label{eqn:acc}\\
  &\frac{tp}{tp+fp} &&\text{precision}\label{eqn:precision}\\
  &\frac{tp}{tp+fn} &&\text{recall}\label{eqn:recal}\\
  & \frac{precision\cdot recall}{(precision+recall)/2} && \text{F1 score}\label{eqn:f1}\\
  % &\frac{2\times tp}{2\times tp+fn+fp} && \text{Dice}\label{eqn:dice_score}
\end{align}

% For multi-class problems, the accuracy is defined as the ratio of predictions that match the ground truth over all predictions. 

tp stands for true positive, fp is false positive and fn is false negative. We use macro $F_1$ score in reporting classification and segmentation results. Macro $F_1$ assigns the same weight to each class-wise score prior to averaging over all classes to take the class imbalance present in most of our validation datasets into account.

We use the $L_1$ error for the regression task. $L_1$ is defined as the mean absolute difference (MAD) between ground truth ($gt_i$) and the prediction (pred) over $N$ samples, or $\sum_{i=1}^N |gt_i-pred_i|$. The subscript $i$ indexes the ground truth and prediction belonging to sample $i$. % Similarly, mean squared error (MSE) is $\sum_{i=1}^N (gt_i-pred_i)^2$.

\section{Foreground filtering}\label{apx:Foreground filtering}

Due to their immense size, WSIs contain many regions that may be irrelevant for histopathology, such as ink, creases, fat, and white background. We threshold each WSI in HSV color space, with the following threshold parameters: $0.65>hue>0.5$, $saturation>0.1$, and $0.9>value>0.5$. The bounds are experimentally determined using a public breast cancer segmentation dataset \citep{aresta2019bach} to isolate the irrelevant background regions. Each WSI is tiled into patches of size $224\times224$ pixels, and patches with foreground ratios $\geq$ 50\% are used as unsupervised training data.

\section{Hyperparameter and suitable augmentation selection}\label{apx:hyperparam_selection}

\begin{table*}[h]
\centering
    \begin{tabular}{ c| c| c c c c c c}
    Optimizer & Temperature & BACH & NCT & BreakHis & Lymph & Gleason2019 & Average \\
    \midrule
      Lars & 0.05 & 77.2 & 87.2 & 73.9 & 73.2 & 49.8 & 72.2 \\
       & 0.10 & 80.3 & 82.7 & 73.9 & 80.9 & 51.7 & 73.9 \\
       & 0.50 & 73.9 & 81.3 & 75.5 & 69.2 & 51.9 & 70.4 \\
     \midrule
      Lamb & 0.05 & 74.7 & 83.5 & 67.8 & 74.1 & 53.7 & 70.8 \\
       & 0.10 & 79.6 & 84.1 & 71.4 & 71.9 & 51.9 & 71.8 \\
       & 0.50 & 69.7 & 82.7 & 78. & 75.4 & 48. & 70.8 \\
    \bottomrule
    \end{tabular}
   \caption{Comparison between Lars and Lamb optimizers used for large batch pretraining a self-supervised network. In optimizer selection experiments, we use baseline augmentations which consist of $\in \{0^{\circ}, 90^{\circ}, 180^{\circ}, 270^{\circ}\}$ image rotations, flips in each axis with 50\% probability, Gaussian blurring, and light color jittering. Results are shown in macro $F_1$ scores.}\label{tab:hyperparam_sel_optimizer}
\end{table*}

% hyperparam dataset is: train-dataset_48400_6_lmdb
\paragraph{Setup} We conduct experiments to determine the best hyperparameter setting for histopathology images. We pretrain a Resnet 18 using $\sim$ 40 thousand images sampled from 57 unlabeled datasets. We use the average macro F1 metric on five validation datasets for the classification task to select each hyperparameter. We compare temperatures \{0.05, 0.1, 0.5, 1, 2, 10\} and batch sizes \{128, 256, 512\}. We compare Adam \citep{kingma2014adam}, Lars \citep{You2017LargeBT}, and Lamb \citep{you2019large} optimizers to account for the large batch size. We use learning rates $0.3\times batchsize/256$, $\frac{4}{2^{(3-(\log_2(batch size)-9)/2)} \times 100}$, and $10^{-4}$ for Lars, Lamb, and Adam, respectively (as proposed by \cite{chen2020simple} for Lars and by \cite{You2017LargeBT} for Lamb). We use $10^{-6}$ as the weight decay. 

Since the original set of augmentations were applied on natural-scene images, they may not be suitable for histopathology. We conduct ablation studies on data augmentations with minor modifications. For instance, the original paper uses 8\% to 100\% randomly resized crops, whereas we compare minima \{1\%, 5\%, 25\%\}, with 100\% as the maximum in all settings. We train for 1000 epochs per experiment. Since loss is a function of the temperature, performance cannot be assessed by comparing different values of the loss. Therefore each pretrained network is assessed on independent validation datasets. 

\paragraph{Data augmentations} In our experiments, we employ randomly resized crops, 90$^{\circ}$ rotations, horizontal and vertical flips, photometric manipulation (color jittering) and Gaussian blurring as augmentations. We randomly select a rectangle corresponding to 1 to 100\% of the area of the original image patch and resize it to the original image's size, stochastically rotate images with degrees $\in \{0^{\circ}, 90^{\circ}, 180^{\circ}, 270^{\circ}\}$ with equal (25\%) probability, flip in each axis with 50\% probability, and compare color jittering strategies where we adjust (brightness, contrast, saturation, hue) of a given image by a percentage value of the original image. Specifically, we compare three types of jittering augmentations with different strengths: light (brightness=0.4, contrast=0.4, saturation=0.4, hue=0.2), medium (0.8, 0.8, 0.8, 0.2) and heavy (0.8, 0.8, 0.8, 0.4). Each number in parentheses represents the fraction of possible change in the corresponding color property, e.g., $\pm$40\% shift in brightness in the light jittering augmentation setting. A 100\% change in brightness can mean a fully dark image (0\% brightness) or doubling the brightness value (200\%). Finally, we employ Gaussian blurring 50\% of the time, with standard deviation $\sigma \sim \mathcal{U}[0.1, 2.0]$ and kernel size 0.1 $\times$ image patch's side length.

\paragraph{Results} In our initial experiments, we observed Adam did not converge for batch sizes $\geq 256$ and was discarded. A batch size of 512 performed the best for both optimizers for all temperature values and is used in all the subsequent experiments. Lamb is a newer technique where \cite{you2019large} have shown better large batch training performance with faster convergence on convolutional networks. However, we observed Lars outperformed Lamb in most settings and was therefore used for the remaining experiments. We found smaller temperatures generally performed better, and larger ones ({1, 2, 10}) did not converge. The temperature parameter of 0.1 performed the best for both optimizers. For the complete list of experiments, please refer to the Table \ref{tab:hyperparam_sel_optimizer}.

Interestingly, we found that randomly cropping 1\% of the image patch and resizing to its original size outperformed both 5\% and 25\% for medium and heavy color jittering settings. We found color jittering strongly correlated with the quality of learned representations. No color jittering has the lowest training loss since it is easy to distinguish images without tampering with their color properties. However, it was outperformed by both light and heavy jittering settings. Overall, we observed that more aggressive augmentations resulted in better representations. For instance, while 1\% of a $224\times224$ patch (a $2\times2$ square) cannot be used to distinguish the original patch it was extracted from, it still outperformed \{5\%, 25\%\} for the medium jittering setting, which performed the best across all settings by over 0.6\% over the next best setting in $F_1$. We argue that the improvement is due to a regularization effect on the contrastive objective, which prevents saturation and enables the learning of more salient features. The average $F_1$ scores obtained from a combination of different augmentation and random cropping settings are shown in Fig. \ref{tab:hyperparam_sel_augmentations_datasets}.

\begin{figure}[H]
     \centering
     \includegraphics[width=0.4\textwidth]{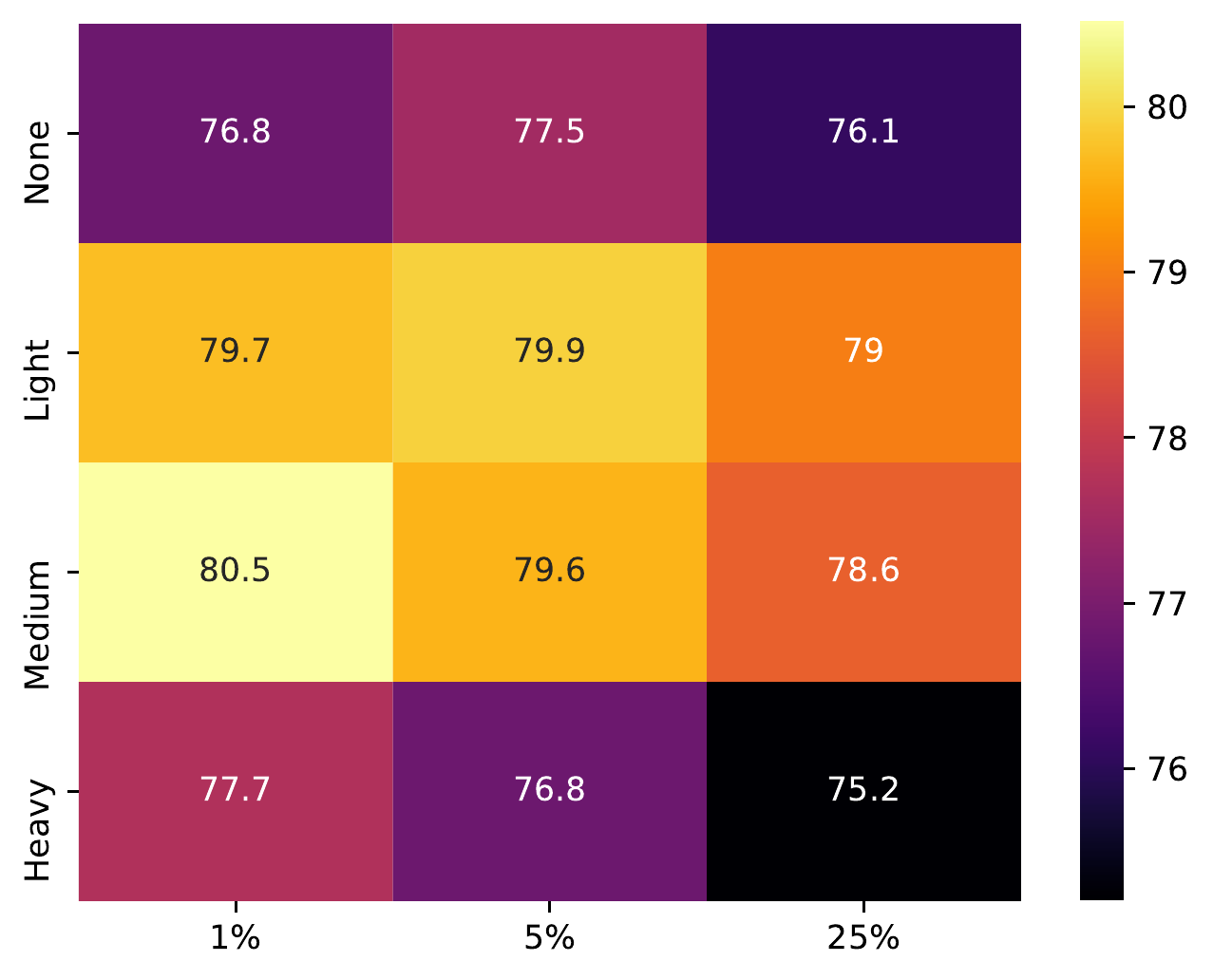}
     \caption{Different combinations of color jittering and random cropping. The figures represent the macro $F_1$ scores averaged over five validation datasets.}
\end{figure}

\subsection{Validation datasets}\label{apx:Validation datasets}

For all validation datasets, training data is comprised of 50\% of total data, with 25\% for both validation and testing sets, unless otherwise mentioned. For classification, we used the following five datasets: BACH challenge dataset \citep{aresta2019bach}, a malignant lymphoma classification dataset which we denote as ``Lymph" \citep{malignant_lymph_dataset}, BreakHis \citep{spanhol2016dataset}, NCT-CRC-HE-100K \citep{katherjakobnikolas20181214456}, and Gleason2019 \citep{gleason2019}. 

The BACH challenge classification dataset contains 400 patches of size $2048\times1536$ pixels extracted from breast biopsy WSIs. The dataset is evenly split into four classes (normal, benign, in situ carcinoma, and invasive carcinoma), and the challenge task was to automatically classify images into these classes. The Lymph dataset contains 374 H\&E stained images from lymph nodes, sampled by multiple pathologists at different sites. The images are separated into three classes: CLL (chronic lymphocytic leukemia), FL (follicular lymphoma), and MCL (mantle cell lymphoma). Notably, this dataset contains a large degree of staining variation within dataset. The BreakHis dataset contains 7909 patches of size $700\times460$ pixels taken from WSIs of breast tumor tissue. The data is labelled as either benign or malignant, and images belong to one of four magnifying factors (40x, 100x, 200x, and 400x). The NCT-CRC-HE-100K dataset contains 107180 patches of size $224\times224$ pixels. The split differs for this dataset, with 75000 patches in the training set, 25000 in validation, and 7180 in the test set. The test set comes from a related dataset called CRC-VAL-HE-7K, which was recommended for use as validation for the larger dataset. Images are taken from 86 stained colorectal cancer tissue slides. The data is split into nine classes: Adipose (ADI), background (BACK), debris (DEB), lymphocytes (LYM), mucus (MUC), smooth muscle (MUS), normal colon mucosa (NORM), cancer-associated stroma (STR), and colorectal adenocarcinoma epithelium (TUM). The Gleason2019 dataset contains 244 tissue micro-array (TMA) images from prostate biopsies. The segmentation labels for this dataset are the Gleason scores for each pixel in the image, ranging from 1 (healthy) to 5 (abnormal).

For segmentation, we used the following two datasets: BACH challenge dataset \citep{aresta2019bach}, and DigestPath2019 \citep{li2019signet}. The BACH segmentation dataset originates from part B of the challenge, and consists of 10 breast biopsy WSIs with pixel-level segmentation masks. These whole-slides are split 5/3/2 for train/validation/test respectively. The segmentation labels are the same 4 classes in the classification dataset. Patches of size $1024\times1024$ pixels were extracted from each slide with a sliding window. The DigestPath2019 challenge dataset contains 250 $5000\times5000$ pixel image patches taken from colonoscopy tissue slides. For our task, we resize each patch and its corresponding pixel-level label image to $1024\times1024$. The segmentation masks provided for this dataset are divided into two classes (benign or malignant), and the aim of this challenge is to use these segmentation labels to identify early-stage colon tumors.

For regression, we used the BreastPathQ dataset \citep{akbar2019automated}. It consists of a total of 2579 patches with size $512\times512$ pixels extracted from 69 WSIs of Post-NAT-BRCA specimens \citep{peikari2017}. BreastPathQ images are labelled according to the percentage of cancer cellularity in each patch. The aim is to predict the percentage cellularity given an input image patch.

\onecolumn

\section{Detailed results}\label{apx:detailed_results}

\begin{table}[H]
\centering
\caption{The downstream task performance of networks trained on top of the pretraining, ImageNet initialization, and randomly initialized network. We report $F_1$ scores averaged over five validation datasets for the classification task, mean $L_1$ difference between the ground truth and the predicted cellularity percentage for one dataset for regression, and average two $F_1$ scores for segmentation. Column with the header R indicates the Resnet model, and Reg. stands for regression.}\label{tab_apx:overall_comparison}
\begin{tabular}{cc| c c c c c c|c|c c c}
\multirow{2}{*}{R} & \multirow{2}{*}{Pretraining} & \multicolumn{6}{c|}{Classification} &
    \multicolumn{1}{c|}{Reg.} &
    \multicolumn{3}{c}{Segmentation}\\
  & & BACH & NCT & BreakHis & Lymph & Gleason2019 & Average & Bpq & BACH & Dp19 & Average \\
\hline
 18 & Random & 47.1 &  69.7 &  72.4 &  50.5 &  38.8 &  55.7 & 13.5 & 42.8 & 75.1 & 59.0 \\
  & ImageNet & 73.1 &  79.8 &  73.9 &  59.9 &  46.7 &  66.7 & 9.3 & 60.5 & 86.6 & 73.6 \\
  & Self supervised & 80.0 &  90.9 &  76.3 &  82.5 &  55.3 &  77.0 & 6.6 & 61.9 & 86.0 & 74.0 \\
 \midrule
 34 & Random & 66.3 &  81.9 &  72.8 &  58.6 &  48.9 &  65.7 & 13.7 & 40.7 & 76.8 & 58.8 \\
  & ImageNet & 74.3 &  80.1 &  71.1 &  70.4 &  45.3 & 68.2 & 7.1 & 45.2 & 87.1 & 66.2 \\
  & Self supervised & 82.0 &  91.4 &  80.2 &  84.3 &  51.8 &  77.9 & 5.9 & 47.2 & 86.0 & 66.6 \\
 \midrule
 50 & Random & 49.8 &  61.2 &  77.2 &  45.4 &  42.9 &  55.3 & 13.3 & 44.1 & 73.6 & 58.9 \\
  & ImageNet & 67.3 &  79.9 &  72.2 &  58.5 &  51.0 &  65.8 & 7.0 & 51.9 & 87.4 & 69.7 \\
  & Self supervised & 81.1 &  86.2 &  78.2 &  84.8 &  54.3 &  76.9 & 5.5 & 46.7 & 86.6 & 66.7 \\
 \midrule
 101 & Random & 50.9 &  54.8 &  76.5 &  43.5 &  39.4 &  53.0 & 15.1 & 41.4 & 74.0 & 57.7 \\
  & ImageNet & 70.6 &  80.7 &  76.3 &  56.3 &  50.9 &  67.0 & 6.7 & 50.7 & 87.4 & 69.1 \\
  & Self supervised & 81.8 &  86.7 &  76.7 &  83.9 &  51.6 &  76.2 & 5.3 & 46.0 & 85.8 & 65.9 \\
 \midrule
 \bottomrule
\end{tabular}
\end{table}

\begin{table}[H]
\centering
\caption{The downstream task performance of linear classifiers trained on top ofthe learned features by self supervision, ImageNet initialization, and randomly initialized network. We report $F_1$ scores averaged over five validation datasets for the classification task, mean $L_1$ difference between the ground truth and the predicted cellularity percentage for one dataset for regression, and average two $F_1$ scores for segmentation. Column with the header R indicates the Resnet model.}\label{tab_apx:last_layer}
\begin{tabular}{cc| c c c c c c|c}
\multirow{2}{*}{R} & \multirow{2}{*}{Pretraining} & \multicolumn{6}{c|}{Classification} &
    \multicolumn{1}{c}{Regression}\\
  & & BACH & NCT & BreakHis & Lymph & Gleason2019 & Average & Bpq \\
\hline
 18 & Random & 23.6 &  23.5 &  73.1 &  34.3 &  23.0 &  35.5 & 27.0 \\
  & ImageNet & 33.7 &  40.3 &  57.3 &  46.9 &  27.1 &  41.1 & 14.9 \\
  & Self supervised & 64.5 &  87.0 &  80.6 &  63.5 &  51.2 &  69.3 & 11.6 \\
 \midrule
 34 & Random & 32.1 &  12.1 &  59.6 &  47.1 &  24.0 &  35.0 & 29.3 \\
  & ImageNet & 39.3 &  47.1 &  63.5 &  41.5 &  31.2 &  44.5 & 15.4 \\
  & Self supervised & 70.6 &  84.5 &  80.0 &  56.4 &  46.6 &  67.6 & 13.5 \\
 \midrule
 50 & Random & 29.1 &  27.7 &  68.7 &  23.1 &  22.8 &  34.3 & 29.1 \\
  & ImageNet & 43.1 &  48.3 &  70.0 &  38.4 &  32.4 &  46.4 & 12.0 \\
  & Self supervised & 65.4 &  86.4 &  79.1 &  67.3 &  50.0 &  69.6 & 10.3 \\
 \midrule
 101 & Random & 21.0 &  6.8 &  58.8 &  26.3 &  22.8 &  27.2 & 2490.7 \\
  & ImageNet & 41.3 &  50.8 &  63.9 &  38.8 &  30.4 &  45.0 & 12.9 \\
  & Self supervised & 70.6 &  86.5 &  79.5 &  68.6 &  50.3 &  71.1 &  10.6 \\
 \midrule
 \bottomrule
\end{tabular}
\end{table}

\begin{table}[H]
\centering
\caption{Examining the impact of resolution on task performance. We use images from one resolution for pretraining and evaluate the pretrained network on all available validation datasets. $10\times$, $20\times$, and $40\times$ refer to $1\mu m/pixel$, $0.50\mu m/pixel$, and $0.25\mu m/pixel$, respectively. We report $F_1$ scores averaged over five validation datasets for the classification task, mean $L_1$ difference between the ground truth and the predicted cellularity percentage for one dataset for regression, and average two $F_1$ scores for segmentation.}\label{tab_apx:resolution_exps}
\begin{tabular}{c|c c c c c c|c|c c c}
\multirow{2}{*}{Resolution} & \multicolumn{6}{c|}{Classification} &
    \multicolumn{1}{c|}{Regression} &
    \multicolumn{3}{c}{Segmentation}\\
  & BACH & NCT & BreakHis & Lymph & Gleason2019 & Average & Bpq & BACH & Dp19 & Average \\
\hline
 $10\times$ & 62.8 & 69.0 & 78.6 & 50.6 & 43.9 & 61.0 & 10.9 & 37.8 & 75.1 & 56.5 \\
 $20\times$ & 54.6 & 65.7 & 74.2 & 65.4 & 42.7 & 60.5 & 9.9 & 42.4 & 72.8 & 57.6 \\
 $40\times$ & 66.1 & 73.8 & 73.3 & 69.2 & 44.4 & 65.4 & 9.4 & 42.4 & 81.6 & 62 \\
 $10, 20, 40 \times$ & 65.1 & 74.0 & 75.3 & 76.9 & 43.7 & 67. & 9.3 & 42.2 & 83.9 & 63.0 \\
 \bottomrule
\end{tabular}
\end{table}

\begin{table}[H]
\centering
\caption{The effect of the number of images used for pretraining in the downstream task performance for classification, regression, and segmentation tasks. We report $F_1$ scores averaged over five validation datasets for the classification task, mean $L_1$ difference between the ground truth and the predicted cellularity percentage for one dataset for regression, and average two $F_1$ scores for segmentation.}\label{tab_apx:dataset_size}
\begin{tabular}{c|c c c c c c|c|c c c}
\multirow{2}{*}{\% of images} & \multicolumn{6}{c|}{Classification} &
    \multicolumn{1}{c|}{Regression} &
    \multicolumn{3}{c}{Segmentation}\\
  & BACH & NCT & BreakHis & Lymph & Gleason2019 & Average & Bpq & BACH & Dp19 & Average \\
\hline
 0.01\% & 65.1 & 77.6 & 75.4 & 62.8 & 51.6 & 66.5 & 8.5 & 43.4 & 81.9 & 62.7 \\
 0.1\% & 77.7 & 85.0 & 78.6 & 75.2 & 53.0 & 73.9 & 7.8 & 61.0 & 84.5 & 72.8 \\
 1\% & 80.0 & 90.9 & 76.3 & 82.5 & 55.3 & 77.0 & 6.8 & 55.7 & 85.9 & 70.8 \\
 10\% & 87.0 & 89.9 & 75.4 & 90.1 & 53.1 & 79.1 & 6.7 & 60.8 & 86.2 & 73.5 \\
 \bottomrule
\end{tabular}
\end{table}

\begin{table}[H]
\centering
   \caption{Selecting suitable augmentations for self-supervised learning on digital histopathology tasks. Baseline augmentations consist of $\in \{0^{\circ}, 90^{\circ}, 180^{\circ}, 270^{\circ}\}$ image rotations, flips in each axis with 50\% probability, Gaussian blurring, and light color jittering. Results are shown in macro $F_1$ scores.}\label{tab:hyperparam_sel_augmentations_datasets}
    \begin{tabular}{ c| c| c c c c c c}
    Random crop & Augmentation intensity & BACH & NCT & BreakHis & Lymph & Gleason2019 & Average \\
    \midrule
      1\% & None & 75.9 & 89. & 79.5 & 87.1 & 52.3 & 76.8 \\
       & Light & 81.7 & 93.6 & 82.7 & 88. &  52.3 & 79.7 \\
       & Medium & 84.4 & 94.4 & 80.7 & 88.2 & 54.8 & 80.5 \\
       & Heavy & 82.4 & 93.1 & 77.2 & 78.1 & 54.2 & 77.0 \\
    \midrule
      5\% & None & 78. & 88.7 & 78.9 & 87.1 & 54.9 & 77.5 \\
       & Light & 82.4 & 94.2 & 80.8 & 88.1 & 53.8 & 79.9 \\
       & Medium & 85.3 & 93.8 & 81. & 83.9 & 54.2 & 79.6 \\
       & Heavy & 82.5 & 93.2 & 77.4 & 78.4 & 52.6 & 76.8 \\
    \midrule
      25\% & None & 66.5 & 77.1 & 74.7 & 71.3 & 53.5 & 68.6 \\
       & Light & 69.7 & 82.7 & 78. & 75.4 & 48. & 70.8 \\
       & Medium & 73.9 & 81.3 & 75.5 & 69.2 & 51.9 & 70.4 \\
       & Heavy & 81.9 & 92.6 & 75.7 & 73.7 & 52.2 & 75.2 \\
    \bottomrule
    \end{tabular}
\end{table}

\section{Comparison to other self-supervised methods}\label{apx:other_methods}

In this section, we compare SimCLR's performance to the earlier self-supervised techniques. We select each technique based on their recency. Each method uses the 1\% of all available pretraining data. We repurpose the segmentation network used in our experiments \citep{Yakubovskiy:2019} with the Resnet50 encoder, and optimize the image reconstruction objective with $L_1$ loss to train an autoencoder. We use an earlier self supervised technique which is based on predicting the RGB channels of a grayscale image for learning representations (colorization) \citep{zhang2016colorful}, and use a more recent self supervised method called CPCv2 that encodes small patches from a larger image in a sliding window. Then, the aim is to predict encoded representations based on spatial proximity (patches extracted from an image that are spatially adjacent are considered more similar) \citep{henaff2019data}.

\begin{table}[H]
\centering
\caption{Comparison of different self-supervised techniques. Each method uses the Resnet50 architecture as its encoder.}\label{tab_apx:other_method_exps}
\begin{tabular}{c|c c c c c c|c|c c c}
\multirow{2}{*}{Setting} & \multicolumn{6}{c|}{Classification} &
    \multicolumn{1}{c|}{Regression} &
    \multicolumn{3}{c}{Segmentation}\\
  & BACH & NCT & BreakHis & Lymph & Gleason2019 & Average & Bpq & BACH & Dp19 & Average \\
\hline
 Ours & 81.1 &  86.2 &  78.2 &  84.8 &  54.3 & 76.9 & 5.5 & 46.7 & 86.6 & 66.7 \\
 CPCv2 & 74.3 &  80.1 &  71.1 &  70.4 &  45.3 & 68.2 & 7.8 & 41.1 & 75.3 & 58.2 \\
 Colorization & 69.2 & 80.2 & 72.4 & 60.0 &  50.5 & 66.5 & 12.2 & 40.9 & 74.1 & 57.5 \\
 Autoencoder & 29.2 &  37.0 &  36.0 &  26.1 &  22.9 & 30.2 & 11.7 & 28.8 & 52.9 & 40.9 \\
 \bottomrule
\end{tabular}
\end{table}

\section{Datasets}

\subsection{Pretraining datasets}\label{apx:Pretraining datasets}

\begin{table}[H]
    \centering
    \begin{scriptsize}
    \caption{Datasets used for pretraining models.}
    \begin{tabular}{c|c|c|c|c|c}
        \hline
        \multicolumn{6}{c}{Image Patch Datasets} \\
        \hline
        Dataset&Organ&Resolution&Staining&Number of patches&Image Type \\
        \hline
        AML-Cytomorphology LMU \citep{AMLCytomorph} &blood &100x & Wright's stain &18365&patch \\
        andrewjanowczyk epi \citep{andrejanowczyk} &breast &20x & H\&E &125&patch \\
        andrewjanowczyk lymphoma \citep{andrejanowczyk}&various &40x & H\&E &374&patch \\
        andrewjanowczyk mitosis \citep{andrejanowczyk} &breast &40x & H\&E &311&patch \\
        andrewjanowczyk nuclei \citep{andrejanowczyk}&breast &40x & H\&E &142&patch \\
        andrewjanowczyk tubule \citep{andrejanowczyk}&colorectal &40x & H\&E &85&patch \\
        % bach part A &breast &20x & H\&E &400&patch \\
        % BreaKHis v1 &breast &4, 10, 20, 40x & H\&E &7909&patch \\
        % breastpathq &breast &20x & H\&E &47345&patch \\
        C-NMC training data \citep{cnmc2019}&blood, bone &Unknown & H\&E &10661&patch \\
        CoNSeP \citep{consep2019}&colon &40x & H\&E &41&patch \\
        CRCHistoPhenotypes \citep{crchisto2016}&colorectal &20x & H\&E &200&patch \\
        Pannuke Fold 1 \citep{gamper2019pannuke}&various &various & H\&E &2656&patch \\
        glas 2015 \citep{sirinukunwattana2016gland}&colon &20x & H\&E &165&patch \\
        icpr 2014 \citep{icpr2014}&breast &20, 40x & H\&E &3975&patch \\
        icpr 2012 \citep{icpr2012}&breast &40x & H\&E &100&patch \\
        lyon2019 \citep{lyon2019}&breast, colon, prostate &40x &anti CD3, CD8 &441&patch \\
        MiMM SBILab \citep{mimm2018}&bone &100x &Jenner- Giemsa &85&patch \\
        MoNuSeg Training Data \citep{monuseg2020_1}&Various &40x & H\&E &30&patch \\
        % NCT-CRC-HE-100k &liver, colorectal &20x & H\&E &107180&patch \\
        Prostate-MRI \citep{prostatemri}&prostate &Unsure & H\&E &26&patch \\
        ramtab \citep{ramtab2012}&colon &unsure &various fluorescence  &200&patch \\
        SN-AM B-ALL and MM \citep{snam2019}&blood, bone &100x &Jenner-Giemsa &61&patch \\
        TNBC NucleiSegmentation \citep{tnbcsegmentation}&breast &40x & H\&E &50&patch \\
        warwick beta cell dataset \citep{betacell2011}&pancreas &40x &H-DAB &20&patch \\
        IDCGrade \citep{BOLHASANI2020100341}  & breast &10x,20x,40x & H\&E &922&patch \\
        \hline
        \multicolumn{6}{c}{Whole-Slide Datasets} \\
        \hline
        Dataset&Organ&Resolution&Staining&Number of WSIs&Image Type \\
        \hline
        TCGA-BLA &bladder &40x & H\&E &469&WSI \\
        TCGA-BRCA &breast &20, 40x & H\&E &1987&WSI \\
        TCGA-CESC &cervix &40x & H\&E &327&WSI \\
        TCGA-COAD &colon &20, 40x & H\&E &983&WSI \\
        TCGA-ESCA &esophagus &20, 40x & H\&E &260&WSI \\
        TCGA-GBM &brain &20, 40x & H\&E &1193&WSI \\
        TCGA-HNSC &head, neck &20, 40x & H\&E &791&WSI \\
        TCGA-KICH &kidney &20, 40x & H\&E &275&WSI \\
        TCGA-KIRC &renal &20, 40x & H\&E &1656&WSI \\
        TCGA-KIRP &renal &20, 40x & H\&E &475&WSI \\
        TCGA-LGG &brain &40x & H\&E &728&WSI \\
        TCGA-LIHC &liver &20, 40x & H\&E &492&WSI \\
        TCGA-LUAD &chest &20, 40x & H\&E &1067&WSI \\
        TCGA-LUSC &lung &20, 40x & H\&E &1100&WSI \\
        TCGA-OV &ovary &20x & H\&E &1376&WSI \\
        TCGA-PRAD &prostate &20, 40x & H\&E &736&WSI \\
        TCGA-READ &rectum &20, 40x & H\&E &364&WSI \\
        TCGA-SARC &various&20, 40x & H\&E &290&WSI \\
        TCGA-STAD &stomach &20, 40x & H\&E &771&WSI \\
        TCGA-THCA &thyroid &20, 40x & H\&E &639&WSI \\
        TCGA-UCEC &uterus &20, 40x & H\&E &805&WSI \\
        CPTAC-AML &marrow, blood &40x &PAS, Wright &123&WSI \\
        CPTAC-CCRCC &kidney &20x & H\&E &783&WSI \\
        CPTAC-CM &skin &20x & H\&E &402&WSI \\
        CPTAC-GBM &brain &20x & H\&E &508&WSI \\
        CPTAC-HNSCC &head, neck &20x & H\&E &390&WSI \\
        CPTAC-LSCC &lung &20x & H\&E &1073&WSI \\
        CPTAC-LUAD &lung &20x & H\&E &1065&WSI \\
        CPTAC-PDA &pancreas &20x & H\&E &557&WSI \\
        CPTAC-UCEC &uterus &20x & H\&E &887&WSI \\
        % bach wsi &breast &20x & H\&E &30&WSI \\
        camelyon16 \citep{camelyon16}&lymph node &40x & H\&E &400&WSI \\
        tupac16 \citep{tupac2016} &breast &40x & H\&E &821&WSI \\
        camelyon17 \citep{camelyon2017}&lymph node &40x & H\&E &1000&WSI \\
        SLN-Breast \citep{slnbreast2019}&breast &20x & H\&E &130&WSI \\
        \hline
    \end{tabular}
    \label{tab:pretraining_datasets}
    \end{scriptsize}
\end{table}

% num total wsis = 469+1987+327+983+260+1193+791+275+1656+475+728+492+1067+1100+1376+736+364+290+771+639+805+123+783+402+508+390+1073+1065+557+887+30+400+821+1000+130

\end{document}